\documentclass[aps, prd, twocolumn, showpacs, superscriptaddress, groupedaddress]{revtex4-1} 
\usepackage{graphicx}	
\usepackage{amssymb}
\usepackage{dcolumn}
\usepackage{color}
\usepackage{subfigure, rotating, bm, array}
\usepackage[pagebackref=false, colorlinks=true]{hyperref}
\hypersetup{
linkcolor=blue,     % color of internal links
citecolor=blue,     % color of links to bibliography
urlcolor=blue} 
\begin{document}
\title{Shadows and negative precession in non-Kerr spacetime}
\author{Parth Bambhaniya}
\email{grcollapse@gmail.com}
\affiliation{International Center for Cosmology, Charusat University, Anand, GUJ 388421, India}
\author{Dipanjan Dey}
\email{dipanjandey.icc@charusat.ac.in}
\affiliation{International Center for Cosmology, Charusat University, Anand, GUJ 388421, India}
\author{Ashok B. Joshi}
\email{gen.rel.joshi@gmail.com}
\affiliation{International Center for Cosmology, Charusat University, Anand, GUJ 388421,  India}
\author{Divyesh N. Solanki}
\email{divyeshsolanki98@gmail.com}
\affiliation{Sardar Vallabhbhai National Institute of Technology, Surat GUJ 395007,  India}
\author{Pankaj S. Joshi}
\email{psjprovost@charusat.ac.in}
\affiliation{International Center for Cosmology, Charusat University, Anand, GUJ 388421, India}
\author{Aadarsh Mehta}
\email{aadarshmehta6@gmail.com}
\affiliation{PDPIAS, Charusat University, Anand, GUJ 388421, India}
\date{\today}

\begin{abstract}
It is now known that the shadow is not only the property of a black hole, it can also be cast by other compact objects like, naked singularities. However, there exist some novel features of the shadow of the naked singularities which are elaborately discussed in some recent articles. In the earlier literature, it is also shown that a naked singularity may admit negative precession of bound timelike orbits which cannot be seen in Schwarzschild and Kerr black hole spacetimes. This distinguishable behaviour of timelike bound orbit in the presence of the naked singularity along with the novel features of the shadow may be useful to distinguish between a black hole and a naked singularity observationally. However, in this paper, it is shown that deformed Kerr spacetime can allow negative precession of bound timelike orbits, when the central singularity of that spacetime is naked. We also show that negative precession and shadow both can exist simultaneously in deformed Kerr naked singularity spacetime. Therefore, any observational evidence of negative precession of bound orbits, along with the central shadow may indicate the presence of a deformed Kerr naked singularity.

\bigskip
Keywords: Black hole, Naked singularity, Precession, Galactic center. 
\end{abstract}

\maketitle

\section{Introduction}
Recently, large amount of attention is being given to understand the nature and dynamics of the center of a galaxy. It was the 10th of April 2019 when Event-Horizon-Telescope (EHT) group first showed the image of the central compact object of Messier-87 (M87) galaxy \cite{M87}. It is being expected that the EHT group will also show the image of our galaxy center Sagittarius-A* (Sgr-A*) in the coming years. On the other hand, UCLA galactic center group, GRAVITY, SINFONI group are continuously observing the orbital dynamics of `S' stars which are orbiting around Sgr-A* with very small (w.r.t galactic scale) pericenter distances \cite{Akiyama:2019fyp, Eisenhauer:2005cv, center1}. Observing the shape and time period of the `S' star orbits, it is estimated that the mass of the central compact object of our galaxy (i.e. Sgr-A*) is around $4.3\times 10^6 M_\odot$. Since those stars are very close to the Sgr-A*, it is expected that general relativistic effects may be observable in the orbital dynamics of `S' stars. One of the most important general relativistic effect in the orbital dynamics is the precession of the bound orbits. UCLA, GRAVITY, SINFONI collaborations are closely eyeing on the `S' stars' motion in order to get any evidence of such precession.

It is generally believed that the central compact object of every galaxy should be a super massive black hole \cite{Kormendy}. However, till now there exists no solid proof for that claim. Recently, in many articles, it is shown that shadow is not only the property of a black hole, it can also be cast by other exotic compact objects e.g. naked singularities, worm holes, grava star, etc \cite{Shaikh:2019hbm,Gralla:2019xty,Abdikamalov:2019ztb,Yan:2019etp,Vagnozzi:2019apd,Gyulchev:2019tvk,Shaikh:2019fpu,Dey:2013yga,Dey+15,Shaikh:2018lcc,Joshi2020,Paul2020,Dey:2020haf,Dey:2020bgo,abdujabbarov_2015a, atamurotov_2015, abdujabbarov_2017, abdujabbarov_2015b, younsi_2016, papnoi_2014, bambi_2013a, ohgami_2015, stuchlik_2018, stuchlik_2019}. In \cite{Shaikh:2018lcc}, it is shown that a naked singularity spacetime known as first type of Joshi-Malafarina-Narayan (JMN-1) spacetime can cast similar type of shadow what a Schwarzschild black hole can cast. On the other hand, in \cite{Joshi2020,Dey:2020haf,Dey:2020bgo}, we show that a naked singularity can cast shadow even in the absence of a photon sphere and this is true for both nulllike and timelike naked singularities.
There are large amount of literature where the nature of timelike bound orbits in different spacetimes are elaborately discussed \cite{Martinez:2019nor, Eva,Eva1,Eva2,tsirulev,Joshi:2019rdo,Bhattacharya:2017chr,Bambhaniya:2019pbr,Dey:2019fpv,Bam2020,Lin:2021noq,Deng:2020yfm,Deng:2020hxw,Gao:2020wjz,aa4}. In \cite{Bambhaniya:2019pbr}, we elaborately discuss the nature of bound timelike orbits in JMN-1 naked singularity spacetime. We show that bound timelike orbits around JMN-1 naked singularity can precesses in the opposite direction of particle motion (i.e. known as negative precession). On the other hand, the negative precession of bound orbits is not allowed in Schwarzschild and Kerr spacetime \cite{Bam2020}.
In \cite{Dey:2019fpv}, we fit our theoretical results with the available astrometric data of `S2' star and we also predict the future trajectories of the same having negative and positive precession.

 In \cite{Dey:2020haf}, we show that the two physical phenomenon, i.e. negative precession and shadow, cannot be present simultaneously in JMN-1 naked singularity spacetime, unless we consider a spacetime configuration which has central naked singularity. It is shown that the spacetime configuration can cast shadow due to the presence of a thin shell of matter in it and outside the thin shell of matter negative precession of bound orbits are possible. Therefore, in that paper we claim that any observational evidence of the simultaneous presence of the shadow and the negative precession would imply the non-existence of Schwarzschild or Kerr black hole at the center of our galaxy Sgr-A*. 

In this paper, we proceed further to show that the above mentioned two physical phenomenon can be present simultaneously in a deformed Kerr spacetime. There are many papers where deformed Schwarzschild and Kerr black holes and their physical signature are elaborately discussed \cite{Johannsen Psaltis, Bambi:2011ek, Hansen:2013owa,Rezzolla:2014mua,Yagi:2016jml,Yunes:2011we}. The deformed Kerr spacetime can be described by considering solution in modified theory of gravity \cite{Hansen:2013owa, Yunes:2011we}. In this paper, we use the solutions of deformed Schwarzschild and deformed Kerr spacetimes described in \cite{Johannsen Psaltis}. In \cite{aa2}, using that description, authors investigates the shape of the shadow cast by deformed Kerr spacetime. In this paper, we first briefly discuss about the deformation in Schwarzschild and Kerr spacetime shown in \cite{Johannsen Psaltis}, then we show the nature of the precession of timelike bound orbits in deformed Kerr spacetime along with the shadow cast by it. The important result what we get is that in the deformed Kerr spacetime, negative precession of bound orbits and shadow both can be present simultaneously when the central singularity is naked.

This paper is organised as follows. In Sec.~(\ref{defschkerr}), we discuss how one can derive the deformed Kerr spacetime using Janis-Newman algorithm which is also discussed in \cite{Johannsen Psaltis}. In that section, we also review the structure of event horizon in deformed Kerr spacetime \cite{Johannsen Psaltis}, where we show that unlike Kerr spacetime, due to the non-zero value of deformation parameter ($\epsilon$), the central singularity can be naked for the spin  $0<|a|<M$, where $M$ is the mass of the central singularity. In Sec. (\ref{orbitsec}), we investigate the nature of the precession of timelike bound orbits in the deformed Kerr spacetime, where we show that negative precession can be present in the same. In Sec.(\ref{shadowsec}), we discuss the nature of the shadow cast by deformed Kerr spacetime and we also show the shadow shape when this spacetime allows negative precession of timelike bound orbits. In Sec.(\ref{concludesec}), we conclude by discussing the important results of this paper. Throughout this paper, we consider Newton's gravitational constant $G_N$ and light velocity $c$ as unity.
%%%%%%%%%%%%%5
\begin{figure*}
\centering
\subfigure[Event horizon position for $\epsilon = 0$ (black), $\epsilon = 1$ (Red), $\epsilon = -1$ (Blue) with spin $a=0.6$.]
{\includegraphics[width=70mm]{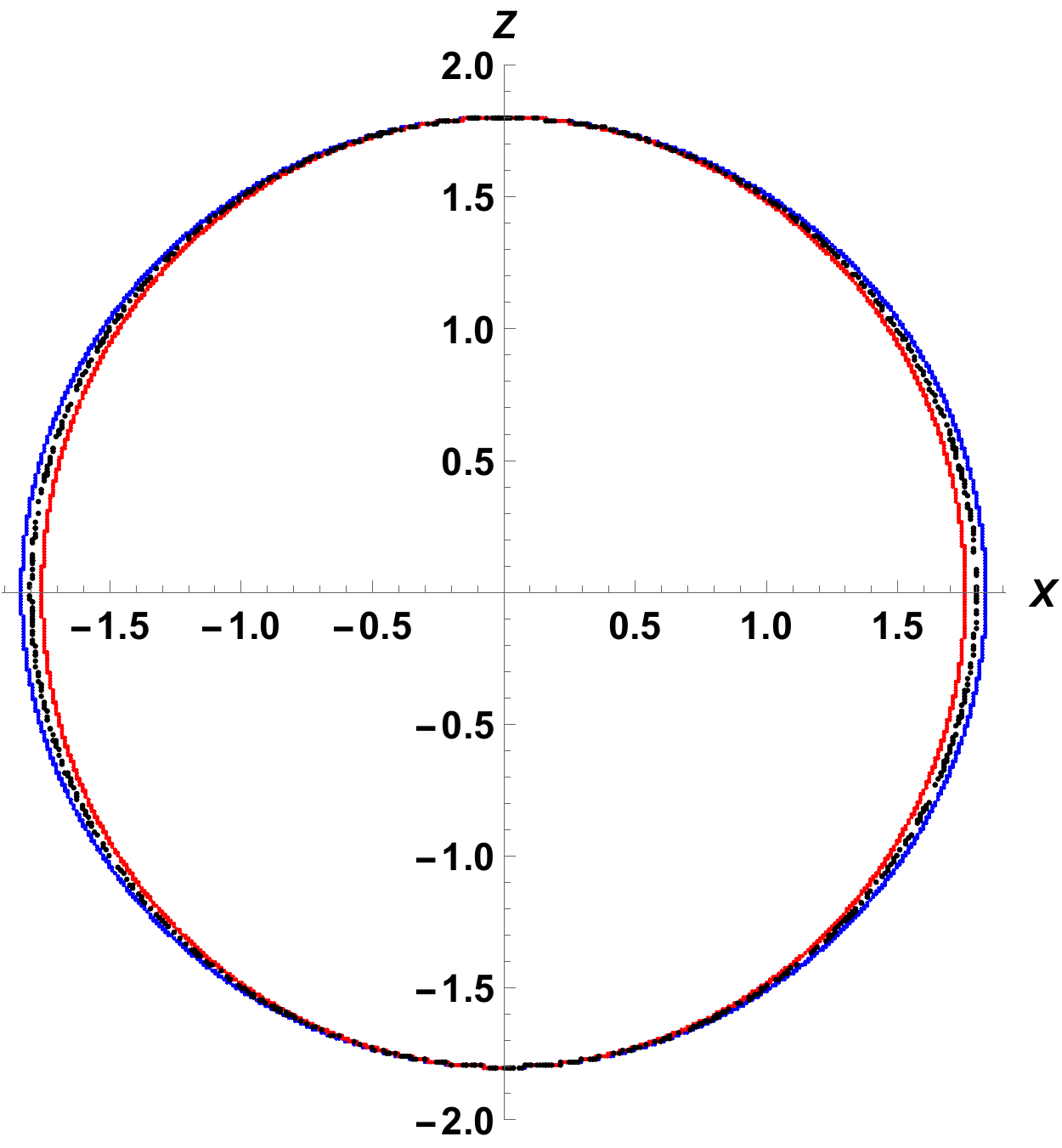}\label{pos1}}
\hspace{0.2cm}
\subfigure[Event horizon position for $\epsilon = 0$ (black), $\epsilon = 1$ (Red), $\epsilon = -1$ (Blue) with spin $a=0.7$.]
{\includegraphics[width=75mm]{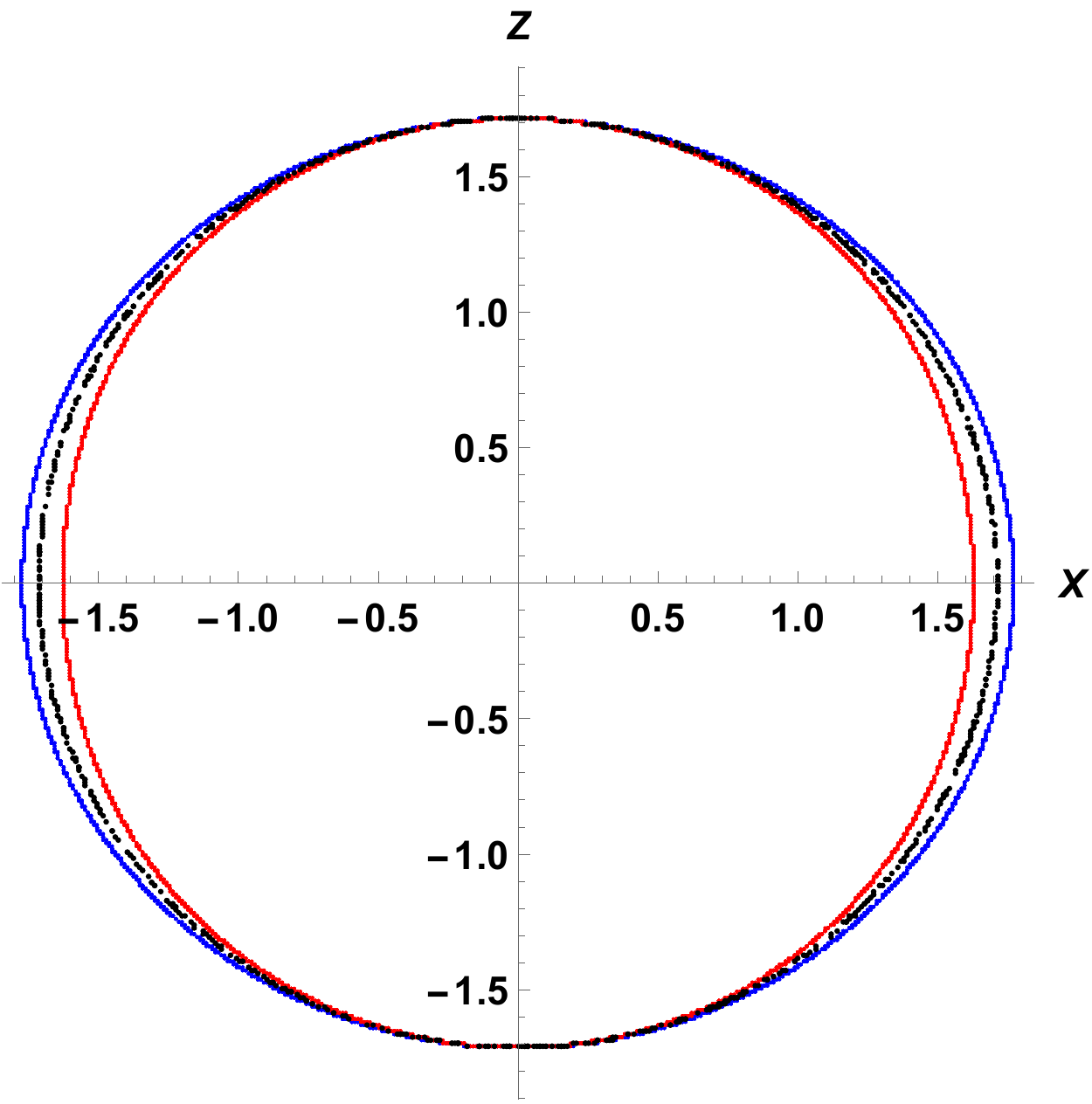}\label{pos2}}
\subfigure[Event horizon position for $\epsilon = 0$ (black), $\epsilon = 1$ (Red), $\epsilon = -1$ (Blue) with spin $a=0.8$. ]
{\includegraphics[width=70mm]{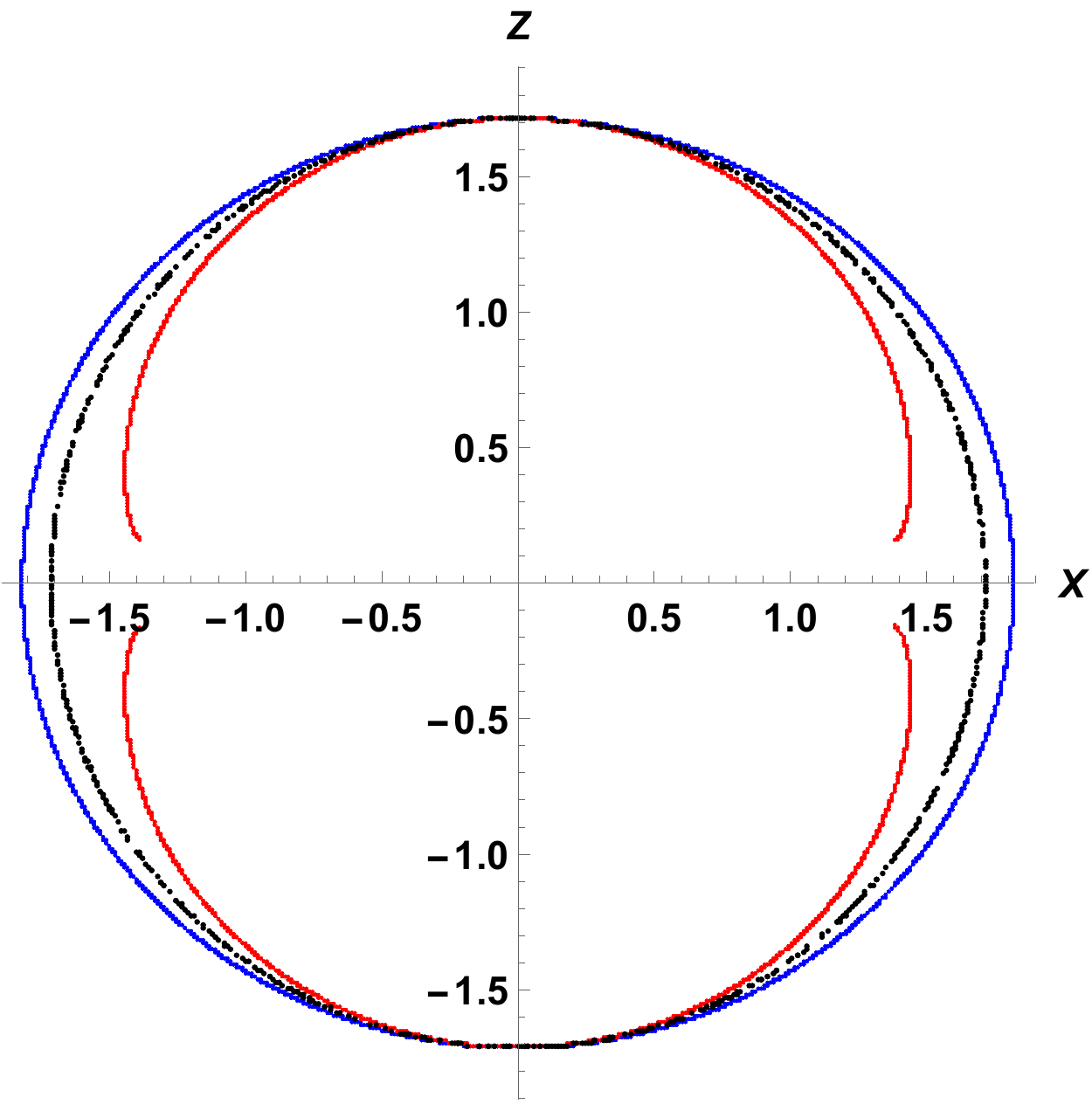}\label{pos3}}
\subfigure[3D view of directional naked singularity for $\epsilon = 1$,  $a = 0.8$. ]
{\includegraphics[width=80mm]{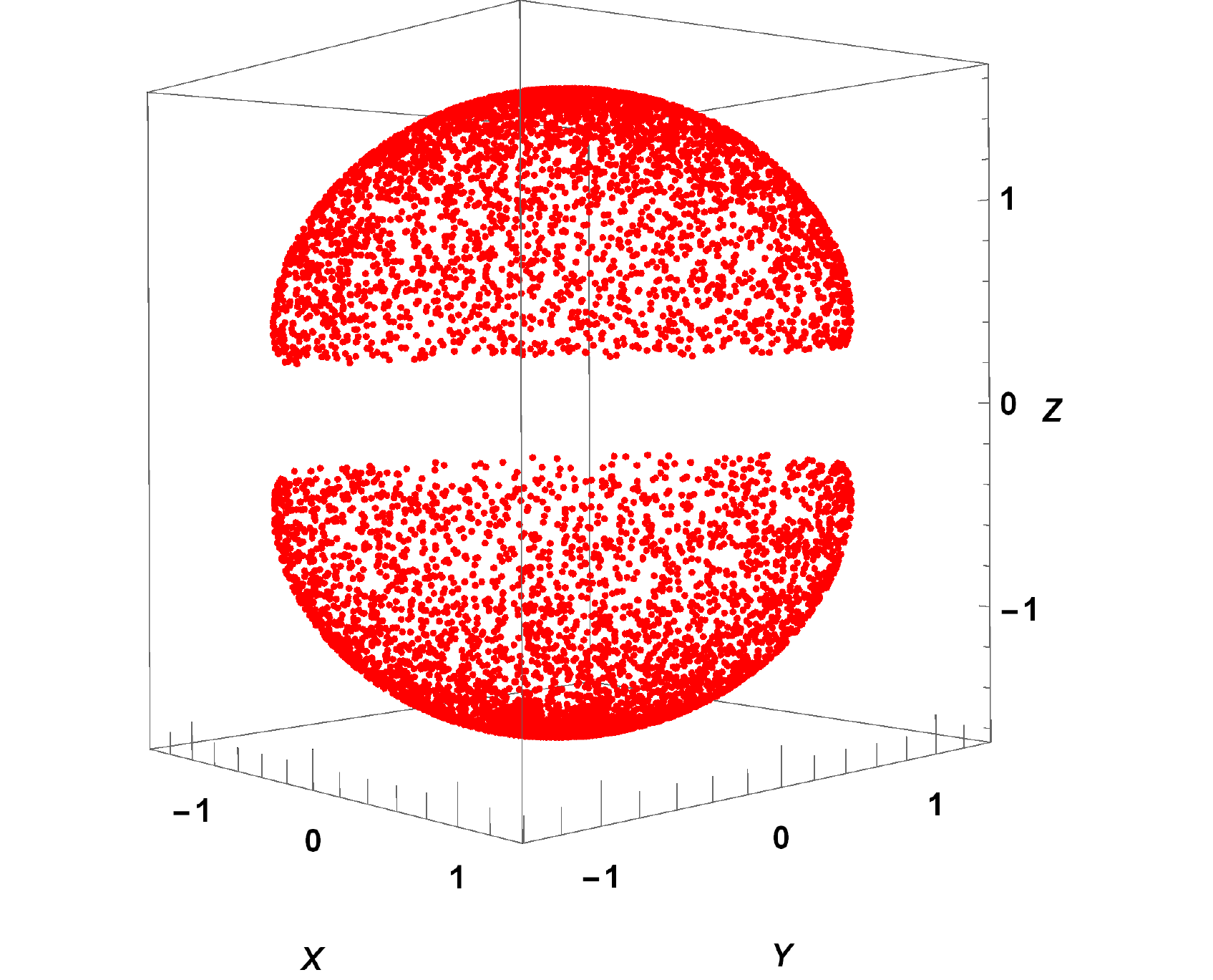}\label{pos3D}}
 \caption{Figure shows the shape and position of event horizon in deformed Kerr spacetime for various values of deformation parameter ($\epsilon$) and spin parameter ($a$). It can be seen that for $\epsilon=1, a=0.8$, there exists directional naked singularity around equatorial plane and the three dimensional view of the directional naked singularity is shown in Fig.~(\ref{pos3D}).}\label{figpos}
\end{figure*}
%%%%%%%%%%%%%%%%

\section{Deformed Schwarzschild and Deformed Kerr black hole}
\label{defschkerr}
The Schwarzschild metric which is a spherically symmetric vacuum solution of Einstein equation, describes the spacetime geometry of a spherically symmetric, static, non-rotating, uncharged black hole. If one introduces the deformations in the Schwarzschild spacetime, the Schwarzschild metric reduces to the following spacetime metric \cite{Johannsen Psaltis}.
\begin{eqnarray}
    ds^2 = -\left(1-\frac{2M}{r}\right)(1+h(r)) dt^2 +\frac{ (1+h(r))}{ \left(1-\frac{2M}{r}\right)}dr^2 \nonumber\\+ r^2 d\Omega^2 \,\, ,
    \label{DSch metric}
\end{eqnarray}
where
\begin{eqnarray}
    d\Omega^2 = d\theta^2 + \sin^2\theta d\phi^2\,\, ,
    h(r) = \frac{\epsilon M^3}{r^3}
\end{eqnarray}
and $M$ is the mass of the deformed Schwarzschild black hole, $h(r)$ is the deformation which describes the deviation from the Schwarzschild metric, and $\epsilon$ is a deformation parameter which is a real constant. Throughout the paper, we consider $M=1$. When $\epsilon=0$ the above metric reduces to the original Schwarzschild metric. In \cite{Yunes:2011we}, it is shown that the above spacetime can be derived as a solution of the Quadratic gravity. In that paper, authors show that a deformation to the Schwarzschild metric solves the modified field equations in the small coupling approximations. They consider a wide class of alternative theories of gravity in four dimensions where in the Einstein-Hilbert action, all the possible quadratic, algebraic curvature scalars multiplied by constant and non-constant couplings are taken.

In 1965, Janis and Newman showed that the Kerr metric could be obtained from the Schwarzschild metric by means of a complex coordinate transformation \cite{Janis Newman}. Same method can be adopted in order to obtain the rotating form of the deformed Schwarzschild metric \cite{Johannsen Psaltis}. Here, one can construct the Kerr-like deformed spacetime metric using the Janis-Newman technique.

In Janis-Newman algorithm, the first step is to consider a seed metric. Here, above mentioned deformed Schwarzschild metric (Eq.~\ref{DSch metric}) is our seed metric.
If we transform the above spacetime metric into the retarded or advanced Eddington-Finkelstein coordinates $(u, r, \theta, \phi)$, then the metric in Eq.~(\ref{DSch metric}) can be written as, 
\begin{equation}
    ds^2 = -\left(1-\frac{2M}{r} \right) (1+h(r)) du^2 - 2(1+h(r)) du dr \\+ r^2 d\Omega^2\,\, ,
    \label{DSch metric in EFC}
\end{equation}
where the coordinate transformation is
\begin{equation}
    du = dt - \left(1-\frac{2M}{r} \right)^{-1} dr\,\, .
\end{equation}
The null tetrads for the above metric (Eq.~\ref{DSch metric in EFC}) can be written as,
\begin{eqnarray}
    l^\mu &=& \delta^{\mu}_{1} \label{l mu}\,\, ,\\
    n^\mu &=& \frac{1}{1+h(r)} \bigg\{\delta^{\mu}_{0} - \frac{1}{2}\bigg(1-\frac{2M}{r} \bigg) \delta^{\mu}_{1} \bigg\} \label{n mu}\,\, ,\\
    m^\mu &=& \frac{1}{\sqrt{2} r} \bigg(\delta^{\mu}_{2} + \frac{i}{\sin\theta} \delta^{\mu}_{3} \bigg) \label{m mu}\,\, ,\\
   \bar{m}^\mu &=& \frac{1}{\sqrt{2} r} \bigg(\delta^{\mu}_{2} - \frac{i}{\sin\theta} \delta^{\mu}_{3} \bigg)\,\, . \label{mbar mu}
\end{eqnarray}
The next step is the complexification of coordinates, in which we allow the coordinates to take complex values. Therefore, the terms in the above expressions which include radial coordinate $r$, can be modified as a real function of $r$ and its complex conjugate $\bar{r}$, for example,
$\frac{1}{r} \to \frac{1}{2}\bigg(\frac{1}{r} + \frac{1}{\bar{r}} \bigg)\, , r^2 \to r\bar{r}$.
Now, a very important step is the complex coordinate transformation.
\begin{eqnarray}
   u^\prime &=& u - ia\cos\theta\,\, ,
     r^\prime = r + ia\cos\theta\,, \,\nonumber\\
     \theta^\prime &=& \theta\,\, ,
     \phi^\prime = \Phi\, .
\end{eqnarray}
Using the above complex coordinate system, the expression of null tetrads shown in Eqs.~ (\ref{l mu}), (\ref{n mu}), (\ref{m mu}), and (\ref{mbar mu}) can be written as,
\begin{widetext}
\begin{eqnarray}
    l^\mu &=& \delta^{\mu}_{1} \label{l new}\,\, ,\\
    n^\mu &=& \frac{1}{1+h(r,\theta)} \bigg\{\delta^{\mu}_{0} - \frac{1}{2}\bigg(1-\frac{2Mr}{\rho^2} \bigg) \delta^{\mu}_{1} \bigg\} \label{n new}\,\, ,\\
    m^\mu &=& \frac{1}{\sqrt{2} (r+ia\cos\theta)} \bigg\{\delta^{\mu}_{2} + \frac{i}{\sin\theta} \delta^{\mu}_{3} + ia\sin\theta \big(\delta^{\mu}_{0} - \delta^{\mu}_{1} \big)\bigg\} \label{m new}\,\, ,\\
    \bar{m}^\mu &=& \frac{1}{\sqrt{2} (r-ia\cos\theta)} \bigg\{\delta^{\mu}_{2} - \frac{i}{\sin\theta} \delta^{\mu}_{3} - ia\sin\theta \big(\delta^{\mu}_{0} - \delta^{\mu}_{1} \big) \bigg\}\,\, , \label{mbar new}
\end{eqnarray}
\end{widetext}
where $\rho^2(r,\theta) = \rho^2 = r^2 + a^2 \cos^2\theta$,
    $ h(r,\theta) = h = \frac{\epsilon M^3 r}{\rho^4}$
Using the above expressions of null tetrads, one can write down the rotating form of the deformed Schwarzschild metric in the null coordinates,
\begin{widetext}
\begin{eqnarray}
    ds^2 = - \bigg(1-\frac{2Mr}{\rho^2} \bigg) \big(1+h(r,\theta) \big) du^2 - 2 \big(1+h(r,\theta) \big) du dr - \frac{4Mar}{\rho^2} \big(1+h(r,\theta) \big) \sin^2\theta du d\Phi\nonumber \\+ 2a\sin^2\theta \big(1+h(r,\theta) \big) dr d\Phi + \rho^2 d\Omega^2 + a^2\sin^4\theta \big(1+h(r,\theta) \big) \bigg(1+\frac{2Mr}{\rho^2} \bigg) d\Phi^2\, .
    \label{DKerr metric in EFC}
\end{eqnarray}
\end{widetext}
The above form of the spacetime metric seems rather complicated. It can be made more simple and symmetrical by transforming it in the Boyer-Lindquist coordinates (BLC). The BLC $(t, r, \theta, \phi)$ are almost always used to write every rotating black hole metrics because all the off-diagonal terms (except $dt d\Phi$) of the metric vanish, and its axial symmetry becomes apparent. The transformation into BLC requires
\begin{figure}
{\includegraphics[width=70mm]{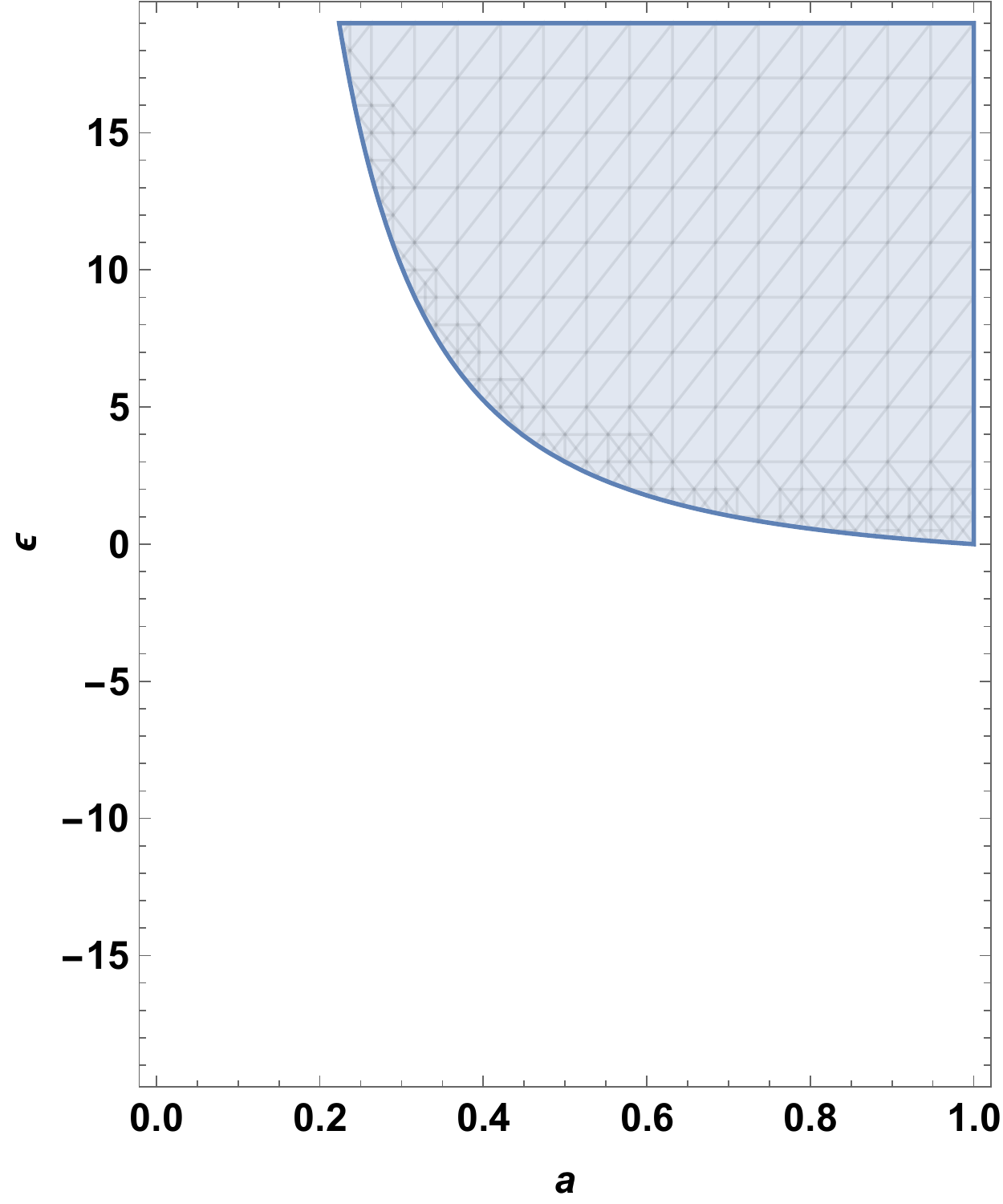}}
\caption{The shaded region in figure shows the existence of naked singularity in deformed Kerr spacetime.}
\label{nakedcond}
\end{figure}
\begin{equation}
du = dt - \xi(r,\theta) dr\, ,
 d\Phi = d\phi - \chi(r,\theta) dr\,\, ,
\end{equation}
where
\begin{eqnarray}
    \xi(r,\theta) &=& 1 + \frac{2Mr}{\Delta + a^2 h \sin^2\theta}\,\, ,\\
    \chi(r,\theta) &=& \frac{a(1+h)}{\Delta + a^2 h \sin^2\theta}\,\, ,\\
    \Delta &=& r^2 - 2Mr + a^2\,\, .
\end{eqnarray}
After the coordinate transformation, we obtain
\begin{eqnarray}
    ds^2 = -\bigg(1-\frac{2Mr}{\rho^2} \bigg) (1+h) dt^2 + \frac{\rho^2 (1+h)}{\Delta + a^2 h \sin^2\theta} dr^2 \nonumber\\+ \rho^2 d\theta^2 - \frac{4Mar}{\rho^2} (1+h) \sin^2\theta dt d\phi\nonumber \\+ \bigg[\rho^2 + a^2\sin^2\theta (1+h) \bigg(1+\frac{2Mr}{\rho^2} \bigg) \bigg] \sin^2\theta d\phi^2\,.
    \label{DK metric}
\end{eqnarray}
In \cite{Johannsen Psaltis}, the above metric is stated as deformed Kerr spacetime, since it can be seen that the above metric reduces to the original Kerr metric when $\epsilon=0$. Kerr spacetime has oblate shape, however, the defomred Kerr spacetime is more oblate for $\epsilon < 0$, and more prolate for $\epsilon > 0$. In \cite{Johannsen Psaltis}, it is stated that the deformation parameter ($\epsilon$) is unconstrained. However, in \cite{aa2}, author discuss the possible maximum value of the deformation parameter ($\epsilon$), where they state that $\epsilon\leq 19.$ Therefore, in this paper, we consider the value of the deformation parameter upto the maximum limit mentioned in \cite{aa2}.

One can derive the position of the inner and outer horizon of deformed Kerr spacetime by solving the following equation \cite{Johannsen Psaltis},
\begin{equation}
g_{t\phi}^2+g_{tt}g_{\phi\phi}=0\,\, .
\end{equation}
In Fig.~(\ref{figpos}), we show the position and shape of outer event horizon of deformed Kerr spacetime for various values of deformation parameter ($\epsilon$) and spin parameter ($a$). Figs.~(\ref{pos1},\ref{pos2},\ref{pos3})show the shapse and position of outer event horizon for spin parameter $a=0.6, 0.7, 0.8$ respectively. It can be seen from Figs.~(\ref{pos3},\ref{pos3D}) that for $ \epsilon=1$ and $a=0.8$, the outer event horizon is not closed and it has a hole near the equitorial plane of deformed Kerr spacetime. Therefore, for that values of $\epsilon$ and $a$, the singularity at the center becomes naked. In Fig.~(\ref{nakedcond}), we show the parameters' space in $\epsilon$ and $a$ frame for which the central singularity of deformed Kerr spacetime becomes naked. The shaded region in that figure shows the possible values of $\epsilon$ and $a$ for the existence of naked singularity in the deformed Kerr spacetime. On the other hand the unshaded region shows the existence of deformed Kerr black hole. From the Fig.~(\ref{nakedcond}), it can be seen that for $0\leq a\leq 1$, the central singularity can be naked when the deformation parameter $\epsilon >0$. However, when $\epsilon<0$ and $0\leq a\leq 1$, the central singularity of the deformed Kerr spacetime cannot be naked.
In the next section, we investigate the nature of precession of timelike bound orbits in the deformed Kerr spacetime.
\section{Timelike Bound Orbits and An approximate solution of the Orbit Equation}
\label{orbitsec}
A general rotating spactime in the Boyer-Lindquist coordinates can be written as,
\begin{equation}
    ds^2=-g_{tt}dt^2+g_{rr}dr^2+g_{\theta\theta}d\theta^2+g_{\phi\phi}d\phi^2-2g_{t\phi}dtd\phi\,\, .
\end{equation}
For the timelike geodesics corresponding to the above spacetime,
the effective potential and the orbit equation are 
\begin{equation}
V_{eff}=\frac{1}{2 g_{rr}} + \frac{e^2-1}{2} + \frac{L^2 g_{tt}+2eL g_{t\phi}-e^2 g_{\phi\phi}}{2 g_{rr}(g_{t\phi}^2+g_{tt}g_{\phi\phi})}
\label{Veff general}
\end{equation}
and
\begin{eqnarray}
\bigg(\frac{dr}{d\phi}\bigg)^2=-\frac{(g_{t\phi})^2+g_{tt}g_{\phi\phi}}{g_{rr}(e g_{t\phi} + L g_{tt})^2} [(g_{t\phi})^2&-&e^2 g_{\phi\phi}+2eL g_{t\phi}\nonumber\\+&g_{tt}&(L^2+g_{\phi\phi})]
\label{general orbit}
\end{eqnarray}
%%%%%%%%%%%%%%%%%%%%%%%%%%%%%%%%%%%%%%%%%%%%%%%%
\begin{figure*}
\subfigure[Deformed Kerr, $0<\epsilon<19$, $a=0.8$]
{\includegraphics[width=88mm]{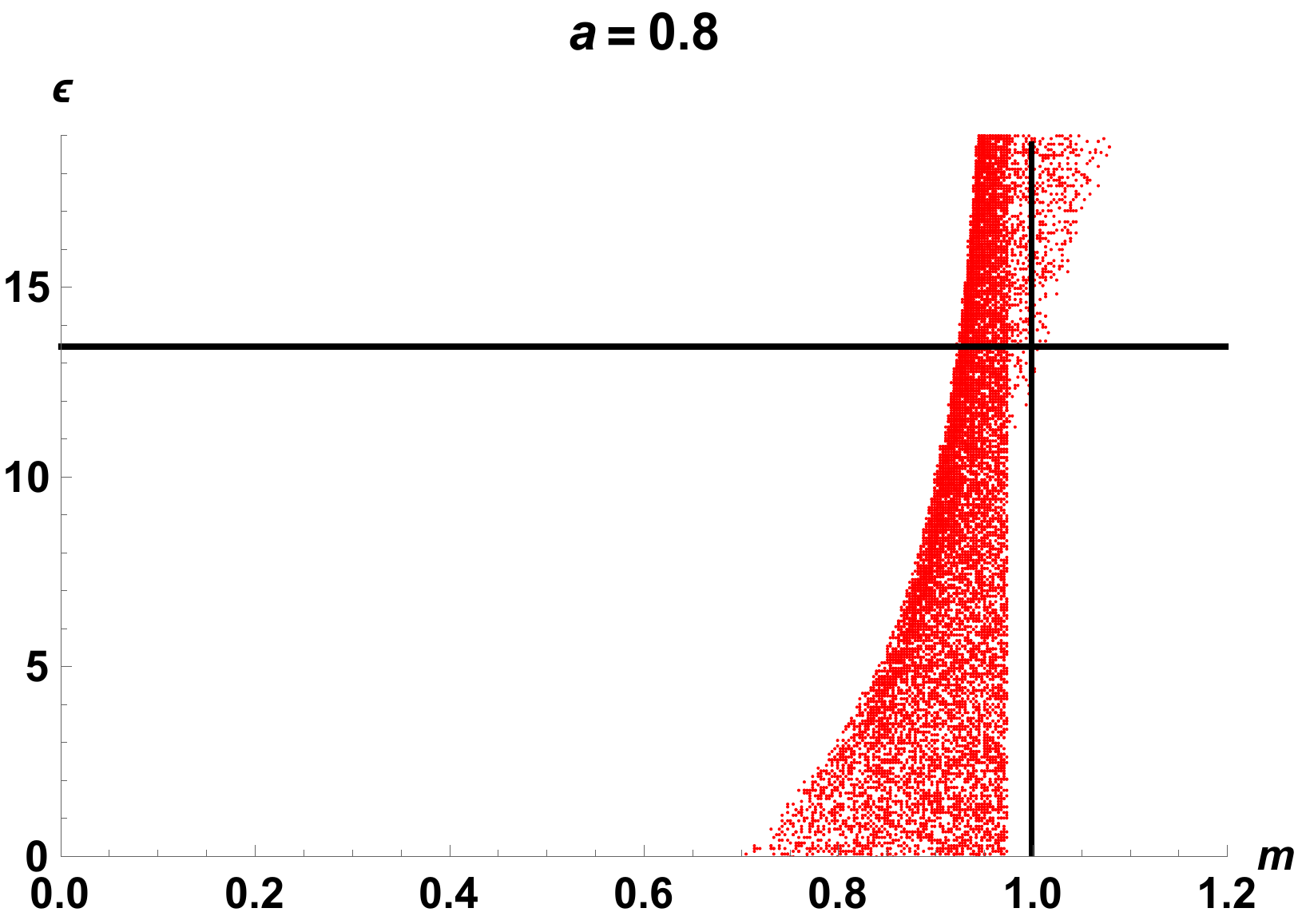}\label{a1}}
\subfigure[Deformed Kerr, $0<\epsilon<19$, $a=0.7$]{\includegraphics[width=88mm]{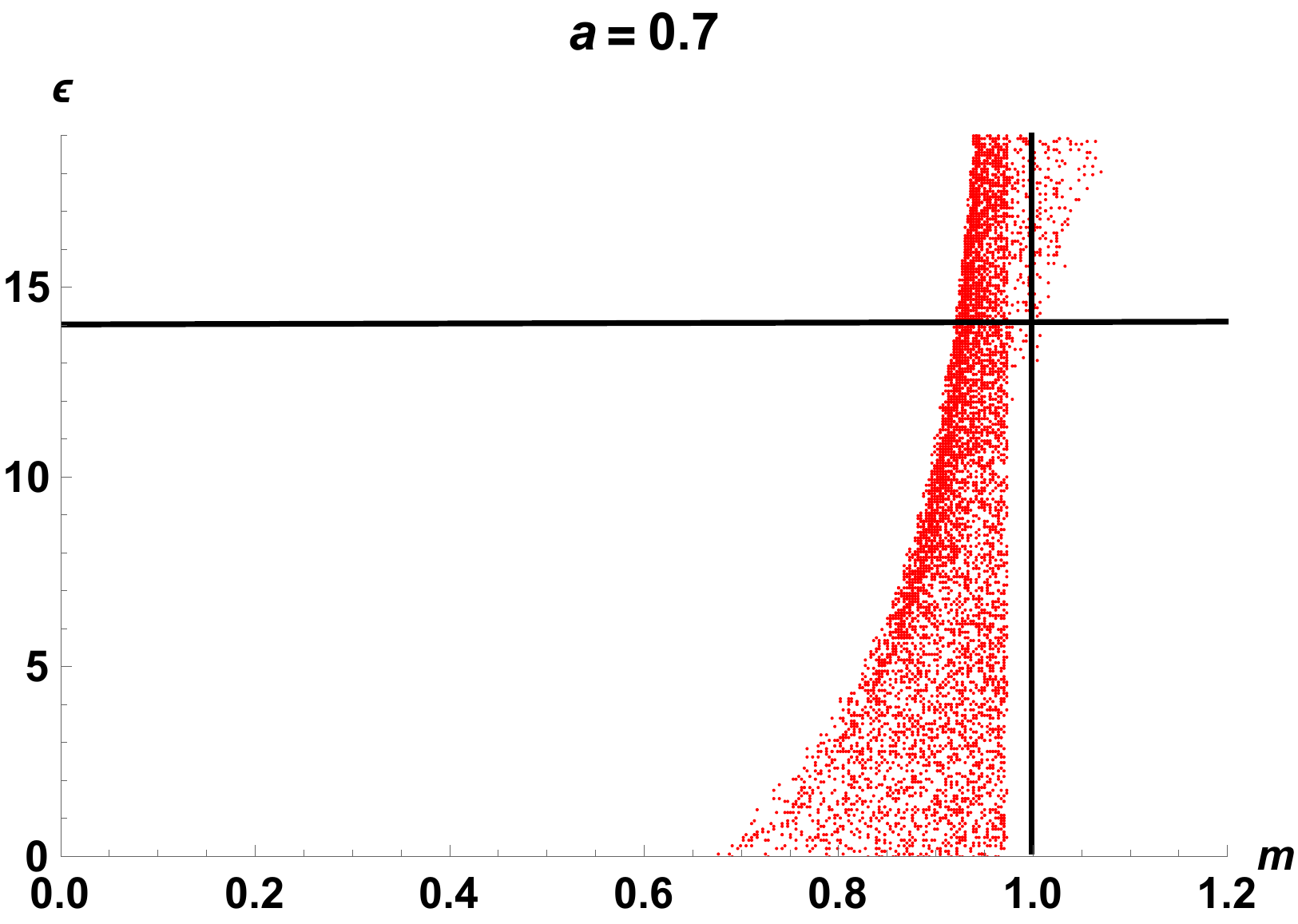}\label{a2}
}
\subfigure[Deformed Kerr, $0<\epsilon<30$, $a=0$]{\includegraphics[width=88mm]{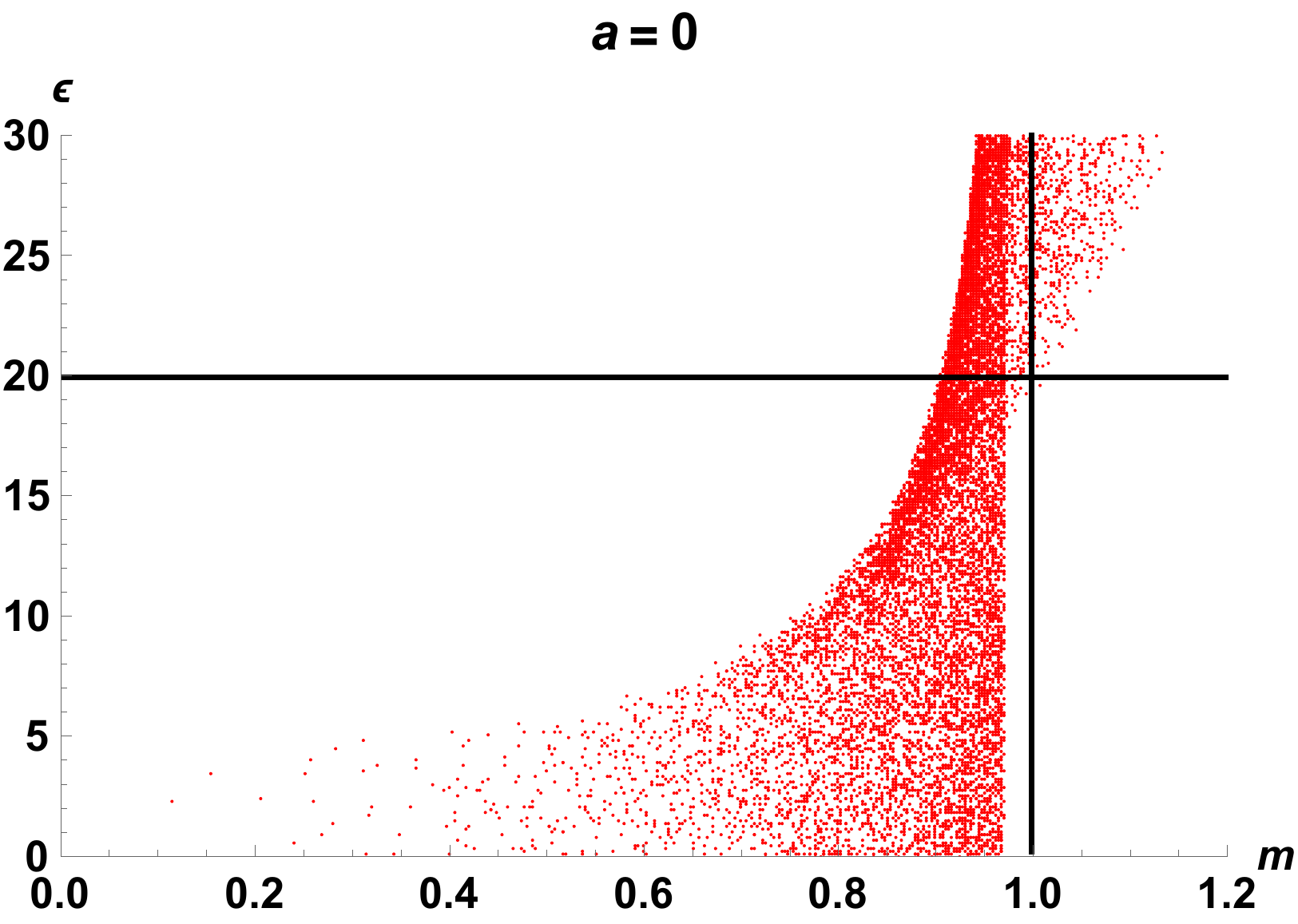}\label{a3}}
\caption{In this figure, we plot the solution points for the timelike bound orbits in deformed Kerr spacetime in $\epsilon$,$m$ frame, where $0<m<1$ and $m>1$ represent positive precession and negative precession of timelike bound orbits respectively. Figs. (\ref{a1}, \ref{a2}, \ref{a3}) show the solution points for the timelike bound orbits in deformed Kerr spacetime for $a=0.8$, $a=0.7$ and $a=0$ respectively.  Here, we consider $M =1$, $h=3$ to $10$, $e=0.965$ to $0.999$.}
\end{figure*}
%%%%%%%%%%%%%%%%%%%%%%%%%%%%%%%%%%%%%%%%%%%%%%%%%%%%%%%%%%%%%%%%%%%%%%%%%%%%%%%%%%%%%%%%%%%%%%%%%%%%%%%%
respectively, where $g_{\mu\nu}$ are the components of the metric tensor, $e$ and $L$ are the conserved energy and the conserved angular momentum per unit rest mass of a test particle. 

Using the deformed Kerr metric in Eq.~(\ref{DK metric}) and the general expression of effective potential of timelike geodesics in Eq.~(\ref{Veff general}), we can write down the following effective potential corresponding to the timelike geodesics in deformed Kerr spacetime,
\begin{eqnarray}
    V_{eff}(r)&=&\frac{1}{2} \bigg\{-1 + e^2 + \frac{a^2}{r^2} - \frac{e^2 r^6}{(r^3 + M^3 \epsilon)^2}    \label{effV}
 \\&-& \frac{a^2 e^2 (2 M + r) - 4 a e L M + (2 M - r) (L^2 + r^2)}{r^3 + M^3 \epsilon}\bigg\}\nonumber\,\, .
\end{eqnarray}
The above equation reduces to the effective potential for timelike geodesics in Kerr spacetime when $\epsilon=0$. As we know, for stable circular timelike geodesic, we need $\frac{dV_{eff}(r)}{dr} = 0$ and $\frac{d^2V_{eff}(r)}{dr^2} > 0$, i.e where the minimum of the effective potential ($V_{min}$) exists. Using the expression of effective potential in Eq.~(\ref{effV}), we can write down $\frac{dV_{eff}(r)}{dr} = 0$ as,
\begin{widetext}
\begin{eqnarray}\label{eq25}
2 M r^{10} + (2 a^2 (-1 + e^2) - 2 L^2)r^9 + (6 (-a e + L)^2 M + 3 (1 - 2 e^2) M^3 \epsilon)r^8 -2 M^4 \epsilon r^7 + (a^2 (-6 + e^2) - L^2) M^3 \epsilon r^6\nonumber \\+ 3 M^4 \epsilon (2 (-a e + L)^2 + M^2 \epsilon) r^5 -4 M^7  \epsilon^2 r^4 + (-a^2 (6 + e^2) + L^2) M^6 \epsilon^2 r^3 -2 a^2 M^9 \epsilon^3 = 0\,\, .
\end{eqnarray}
\end{widetext}
The bound elliptical orbits exist for $V_{min}<E<0$. Where, $E$ is the total energy of a test particle. Using Eq.~(\ref{general orbit}), the orbit equation for the deformed Kerr spacetime can be written as,
\begin{widetext}
\begin{eqnarray}
    \frac{d^2u(\phi)}{d\phi^2} = \frac{1 - 2 M u + a^2 A u^2}{2u A^3 (L + 2 a e M u - 2 L M u)^3} \big[f_6(a,e,L,\epsilon,M,u) u^6 + f_5(a,e,L,\epsilon,M,u) u^5 + f_4(a,e,L,\epsilon,M,u) u^4\nonumber \\+ f_3(a,e,L,\epsilon,M,u) u^3 + f_2(a,e,L,\epsilon,M,u) u^2 + f_1(a,e,L,\epsilon,M,u) u + f_0(a,e,L,\epsilon,M,u) \big]
    \label{orbit_DK}
\end{eqnarray}
\end{widetext}
Where, $u = \frac{1}{r}$, and $A(\epsilon,M,u)=A=1+M^3u^3\epsilon$. The functions used in (\ref{orbit_DK}) are defined in Appendix (\ref{Appendix}). It is very much difficult to solve this second order differential orbit equation analytically. As we have mentioned before, in this section, our primary goal is to investigate the nature of the orbit precession which can be done by solving the above differential equation considering small eccentricity approximation \cite{Bambhaniya:2019pbr, Dey:2019fpv, Bam2020}. One can consider the following expression of $u(\phi)$ as a solution of the differential Eq.~(\ref{orbit_DK}),
\begin{equation}
    u(\phi) = \frac{1}{pM} (1+k\cos(m\phi))\,\, ,
    \label{sol_DK}
\end{equation}
\begin{figure*}
{\includegraphics[width=80mm]{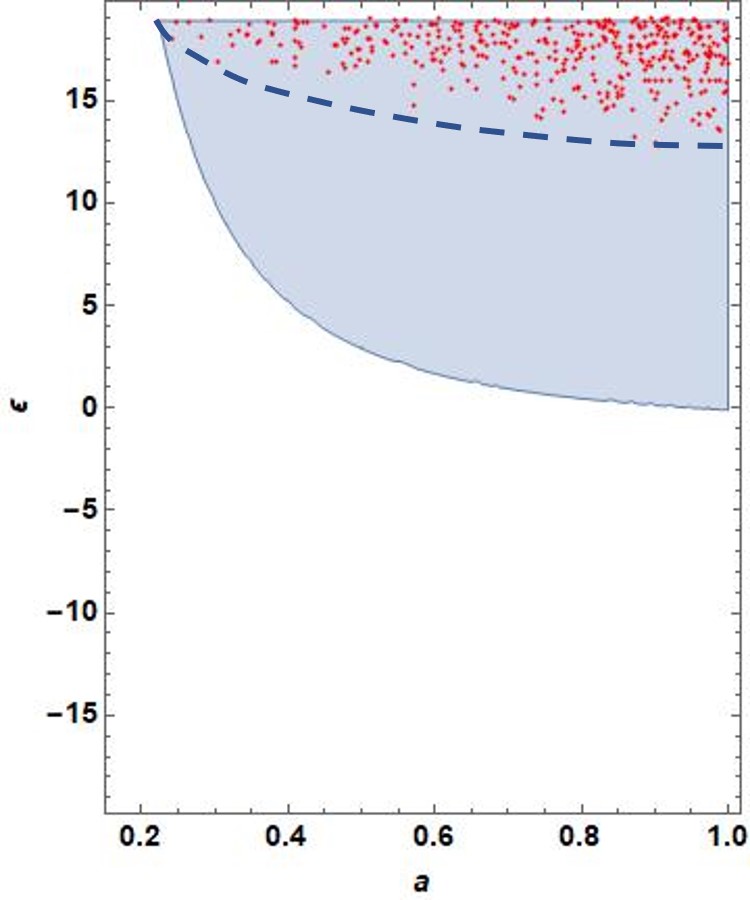}}
\caption{The shaded region in figure shows the existence of naked singularity in deformed Kerr spacetime. The area above the blue dashed line in the shaded region is the region where negative precession is possible, where the red dots are the solution points for negative precession. The remaining unshaded region in $\epsilon$, $a$ frame shows the existence of deformed Kerr black hole.}
\label{enrgcon1}
\end{figure*}
where we consider only the first order term of eccentricity ($k$) and $p$ and $m$ are real positive constants that are used to determine the nature of the orbit precession. The orbit precesses in the direction of particle motion (positive precession) when $m<1$ and the orbit precesses in the opposite direction of particle motion (negative precession) when $m>1$. Substituting Eq.~(\ref{sol_DK}) into the orbit Eq.~(\ref{orbit_DK}), and equating the zeroth order terms and the first order terms from both the sides, one can obtain,
\begin{eqnarray}
    g_4(a,e,L,\epsilon,p) p^8 M^4 + g_2(a,e,L,\epsilon,p) p^4 (p^3 + \epsilon) M^2\nonumber \\+ g_0(a,e,L,\epsilon,p) a^2 (p^3 + \epsilon)^2 = 0
    \label{DK pol}
\end{eqnarray}
And
\begin{widetext}
\begin{eqnarray}
    m^2 = \frac{1}{M^4 p^8 (p^3 + \epsilon)^4 (2 a e + L (-2 + p))^4} \big[ s_6(a,e,L,\epsilon,p) p^{12} M^6 + s_4(a,e,L,\epsilon,p) p^8 (p^3 + \epsilon) M^4 \nonumber\\-3 a^2 p^4 (p^3 + \epsilon)^4 s_2(a,e,L,\epsilon,p) M^2 -a^4 ((p^3 + \epsilon)^4) s_0(a,e,L,\epsilon,p)  \big]\,\, ,
    \label{DK m}
\end{eqnarray}
\end{widetext}
where the functions used in Eqs.~(\ref{DK pol}) and (\ref{DK m}) are defined in Appendix (\ref{Appendix}). The polynomial Eq.~(\ref{DK pol}) can be solved for $p$, and a real positive root can be substituted into the expression of $m$ described in Eq.~(\ref{DK m}). 

In Figs.~(\ref{a1},\ref{a2},\ref{a3}), using the solutions of $m$ for various values of parameters $a,e,L,\epsilon$, we show the solution points in $m$ and $\epsilon$ frame for which bound stable orbits are possible. Note that, the parameters' range are chosen in such a way that there exist solutions of bound orbits where the first order eccentricity approximation is valid \cite{Bambhaniya:2019pbr}. Those three figures show that there exists a critical positive value of $\epsilon$ greater than which negative precession (i.e. $m>1$) of bound timelike geodesics in deformed Kerr spacetime is possible. 
%In Fig.~(\ref{a3}), taking one of the solution points, we show negative precession of bound timelike orbits in deformed Kerr spacetime, where $M = 1, L =  3.20141, a = 0.908584, \epsilon = 18.68, e= 0.988385$. 
In Figs.~(\ref{a1},\ref{a2},\ref{a3}), the values of the spin parameter $a$ are taken as $a= 0.8,~ 0.7,~0$ respectively. It can be seen that for $a=0.8,~0.7$, negative precession is allowed for $\epsilon >13$. However, for $a=0$, negative precession of bound timelike orbits is possible when $\epsilon>20$. As we mentioned previously, there exists an upper limit of $\epsilon$ (i.e. $\epsilon\leq 19$) \cite{aa2}. Therefore, in deformed Schwarzschild spacetime (i.e. $a=0$), negative precession of timelike bound orbits is not possible for the allowed range of $\epsilon$. In Fig.~(\ref{enrgcon1}), we show the allowed region for naked singularity (the shaded region) and the solution points of timelike bound orbits having negative precession (red dots). In that figure, the area above the blue dashed line is the region where the negative precession is allowed. It can be verified that all the solution points of timelike bound orbits having negative precession lies inside the region of naked singularity (i.e. the shaded region). Therefore, it can be concluded that in deformed Kerr spacetime, negative precession can be allowed when the central singularity is naked.
%Fig.~(\ref{a4}) shows the solution points of bound timelike orbits in deformed Schwarzschild spacetime in $m$, $\epsilon$ frame. In that figure, it can be seen that the negative precession of bound orbits in deformed Schwarzschild spacetime is allowed when $\epsilon>0$. Hence, unlike usual Schwarzschild and Kerr spacetime, in deformed Schwarzschild and deformed Kerr spacetime, both the positive and negative precession are allowed. 

In the next section, we show the shape of the shadow cast by deformed Kerr spacetime. As it was mentioned, in Kerr spacetime, shadow and negative precession of bound orbits cannot be present simultaneously. However, from the discussion on shadow in the next section, one can realize the above statement is not valid for deformed Kerr spacetime.

\begin{figure*}
\centering
\subfigure[Shadow of deformed Kerr black hole with $\epsilon = -1$ and $a = 0.6$.]
{\includegraphics[width=55mm]{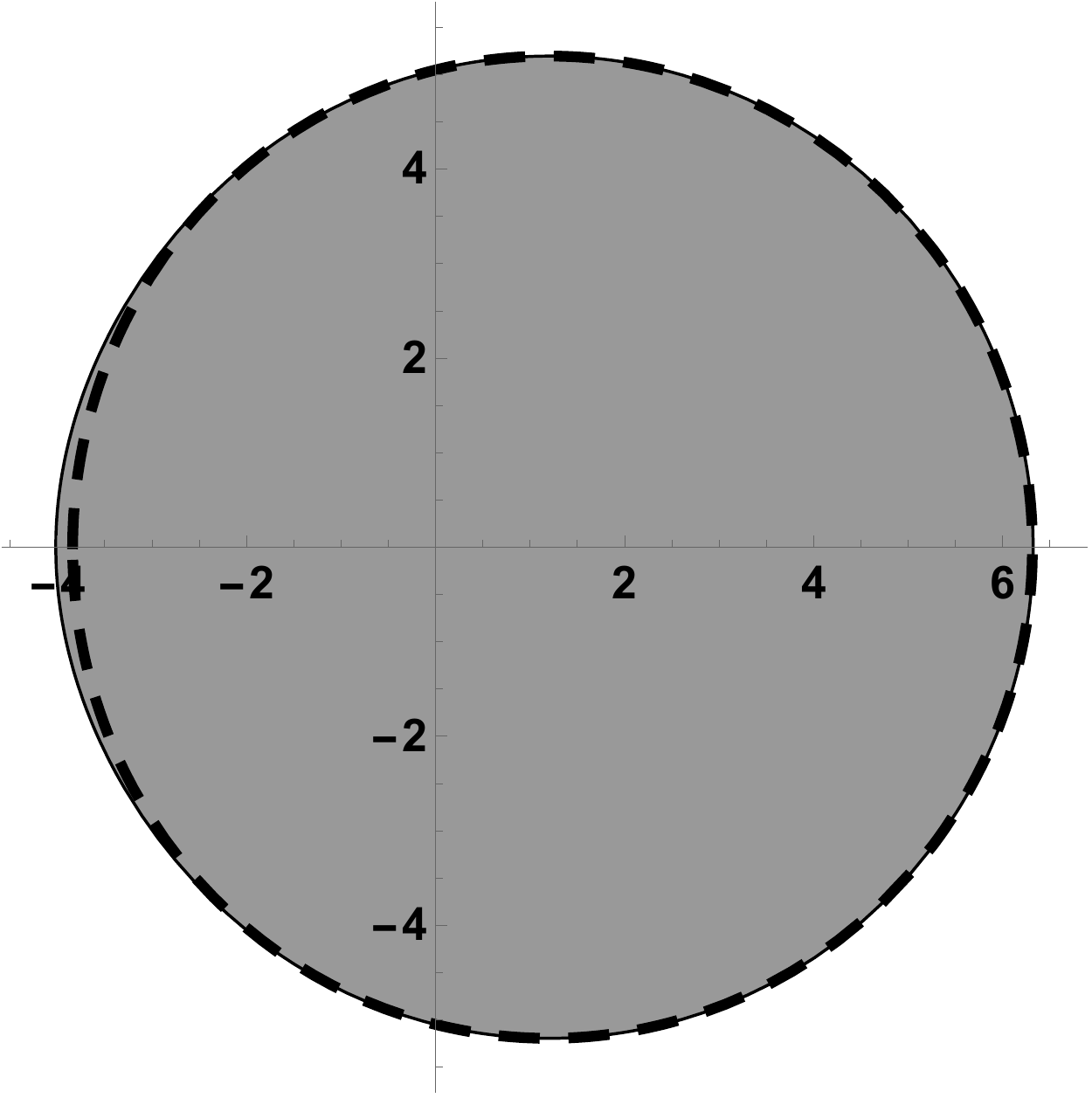}\label{re1n}}
\hspace{0.2cm}
\subfigure[Shadow of deformed Kerr black hole with $\epsilon = -1$ and $a = 0.7$.]
{\includegraphics[width=55mm]{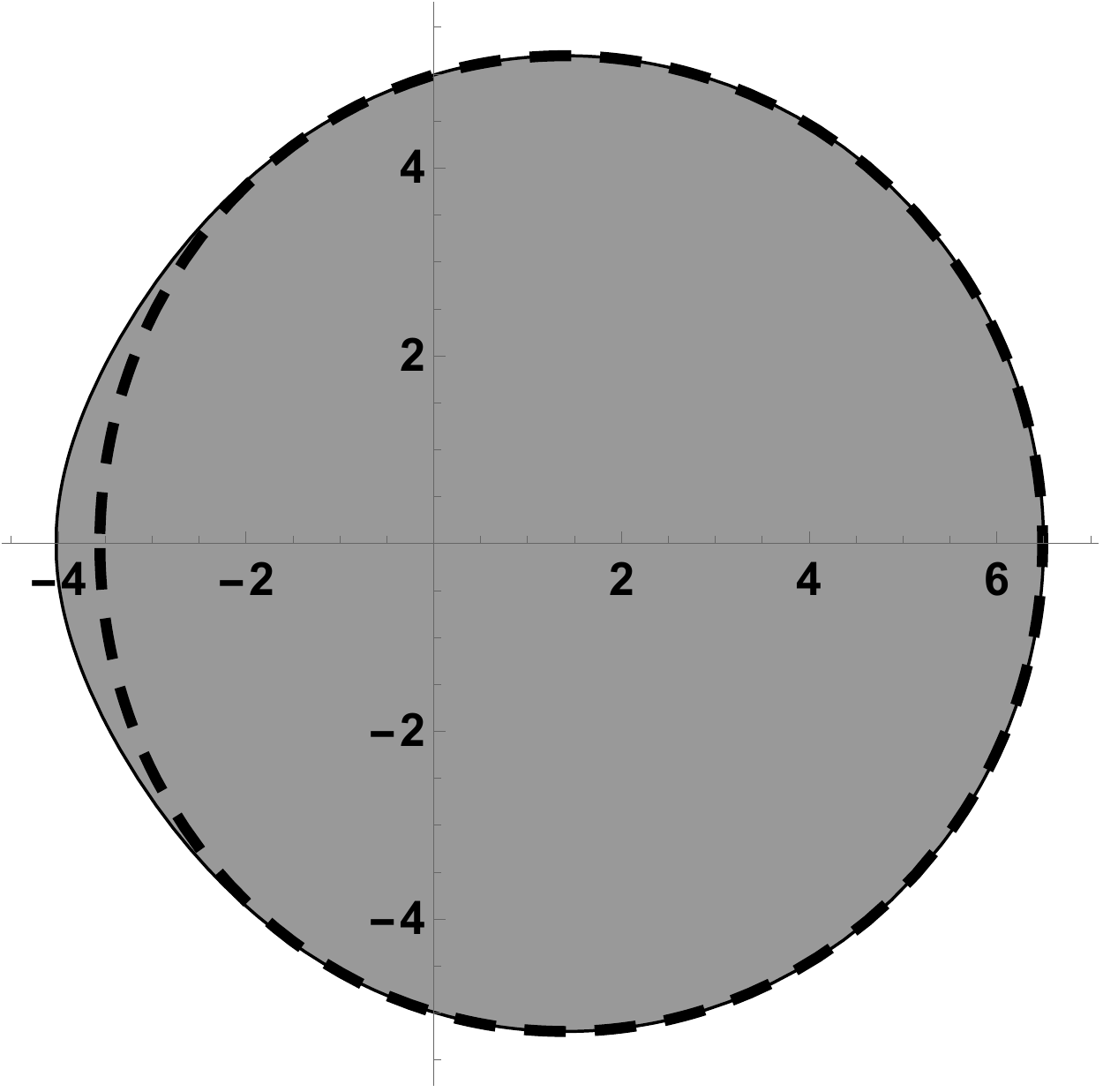}\label{re2n}}
\subfigure[Shadow of deformed Kerr black hole with $\epsilon = -1$ and $a = 0.8$.]
{\includegraphics[width=55mm]{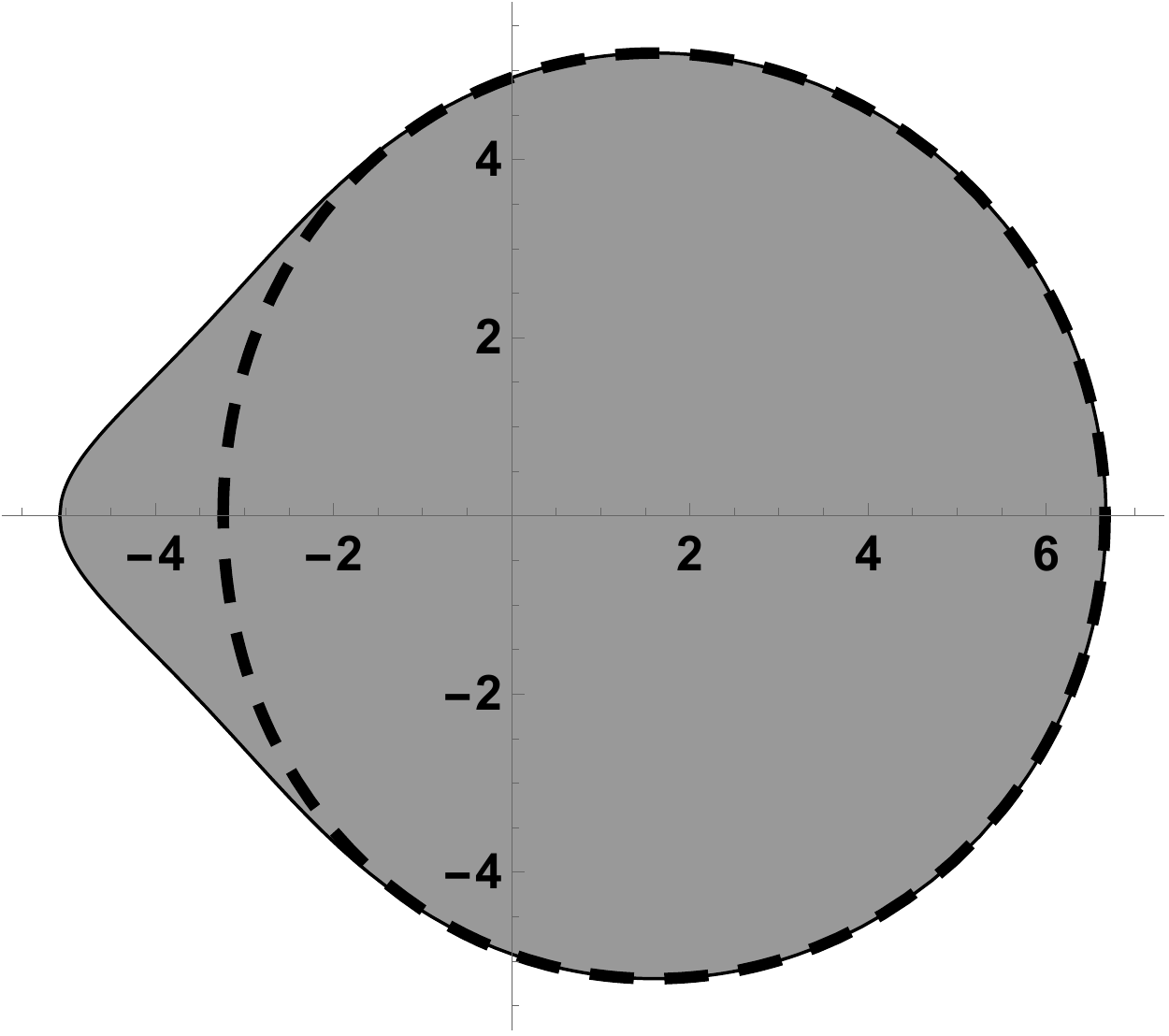}\label{re3n}}\\
\subfigure[Shadow of deformed Kerr black hole with $\epsilon = 1$ and $a = 0.6$.]
{\includegraphics[width=55mm]{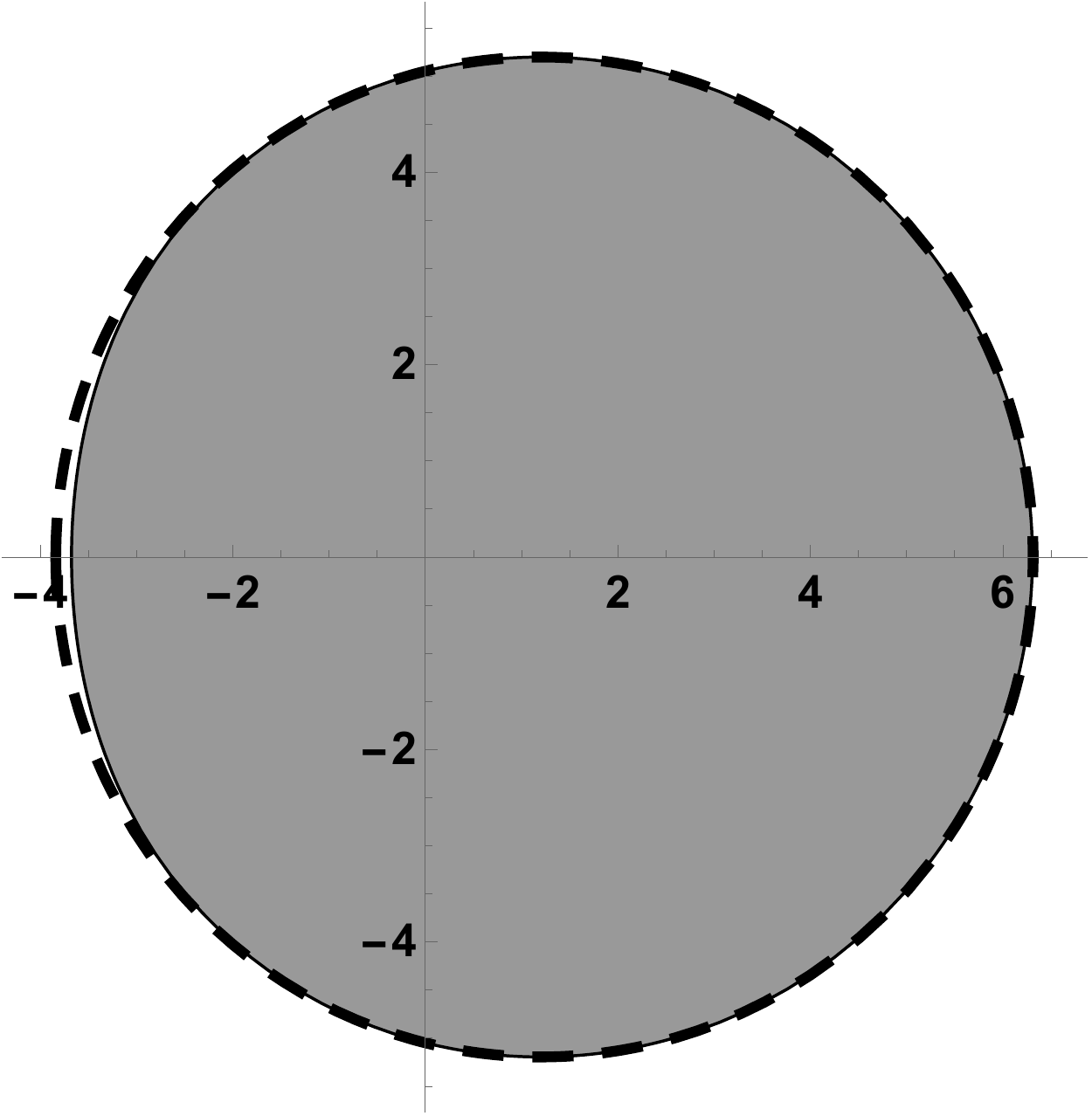}\label{re1p}}
\hspace{0.2cm}
\subfigure[Shadow of deformed Kerr black hole with $\epsilon = 1$ and $a = 0.7$.]
{\includegraphics[width=55mm]{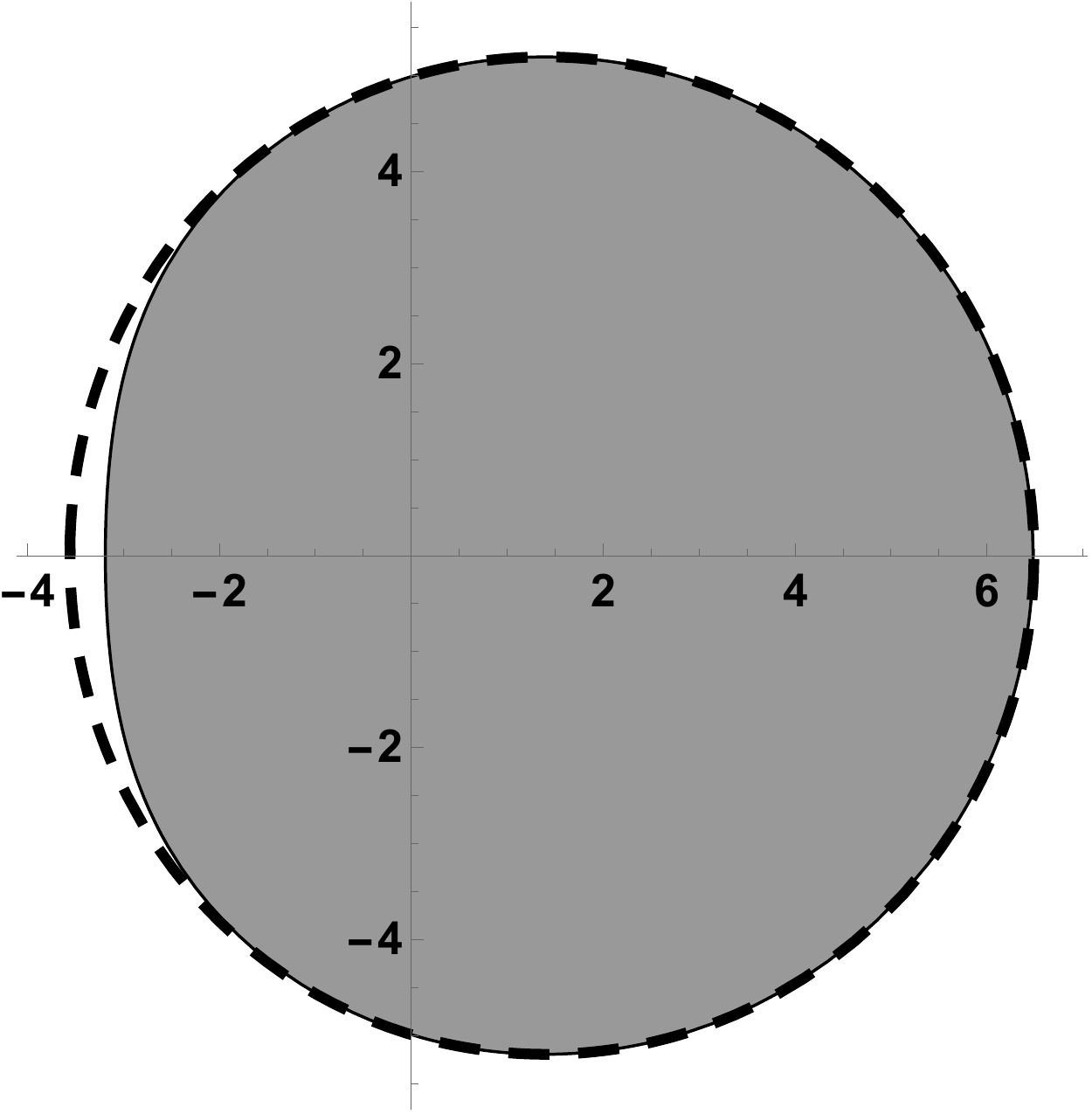}\label{re2p}}
\subfigure[Shadow of deformed Kerr naked singularity with $\epsilon = 1$ and $a = 0.8$.]
{\includegraphics[width=55mm]{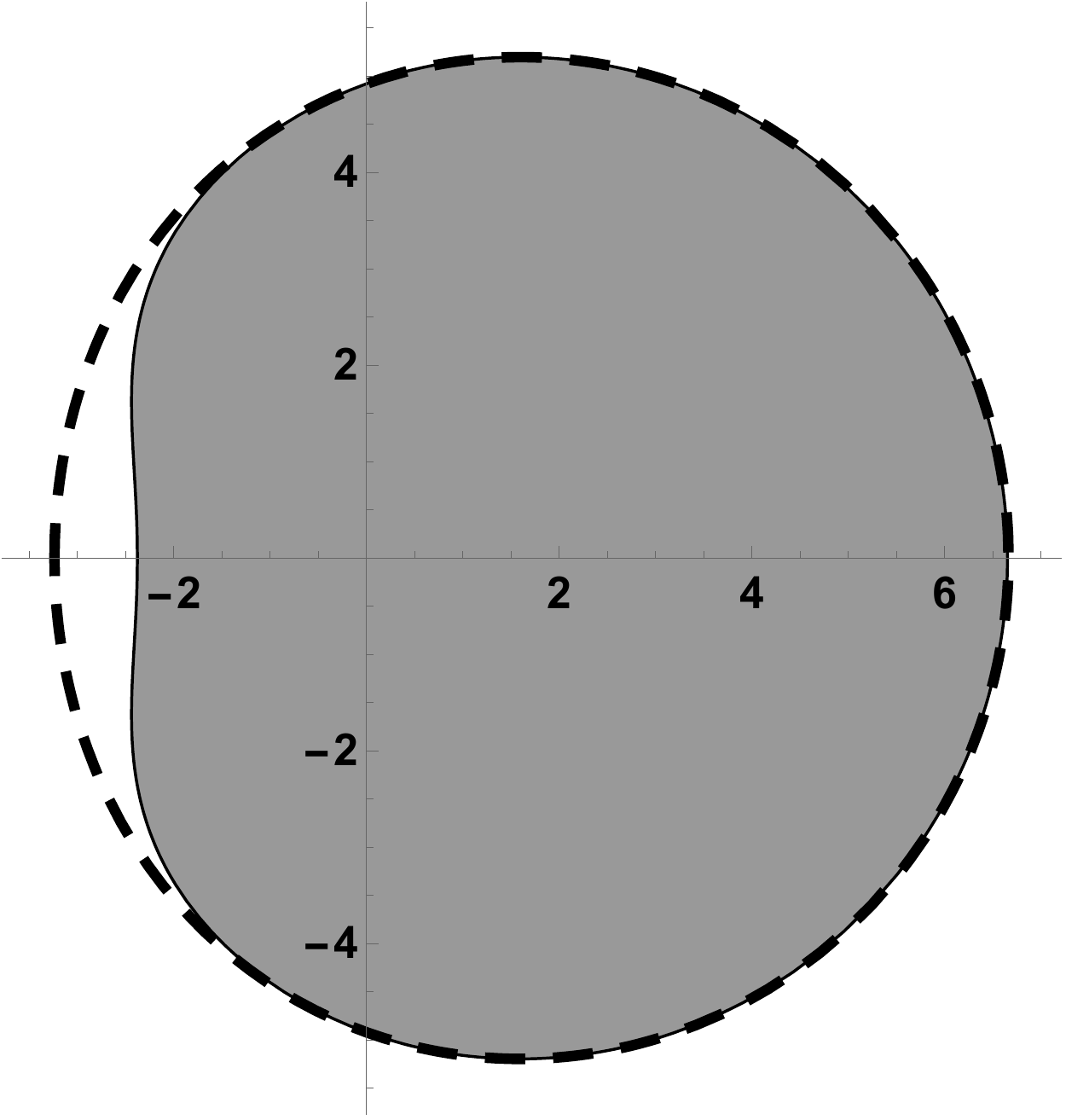}\label{re3p}}\\
\subfigure[Shadow of deformed Kerr naked singularity with $\epsilon = 18$ and $a = 0.6$.]
{\includegraphics[width=55mm]{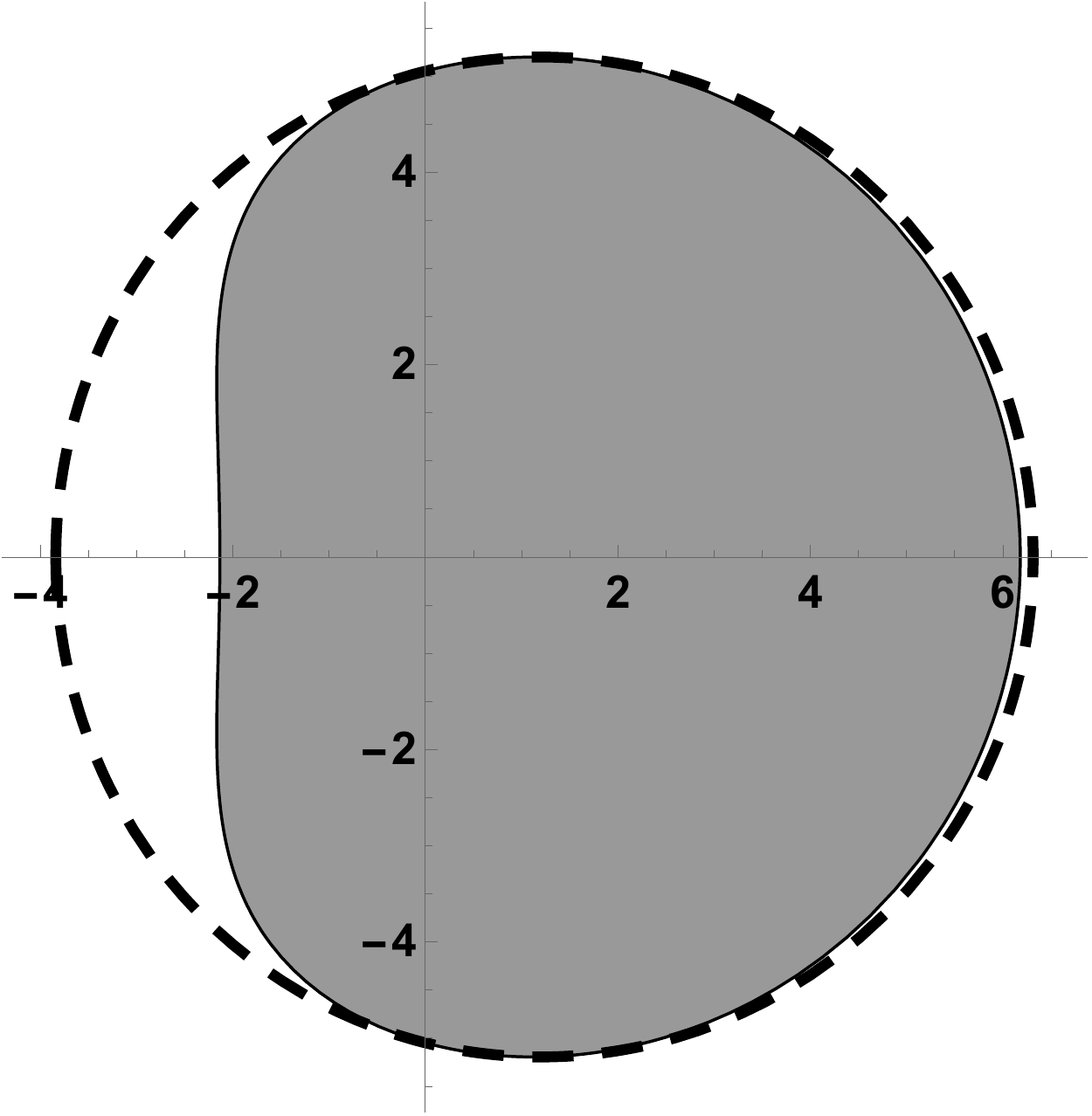}\label{re11p}}
\hspace{0.2cm}
\subfigure[Shadow of deformed Kerr naked singularity with $\epsilon = 18$ and $a = 0.7$.]
{\includegraphics[width=55mm]{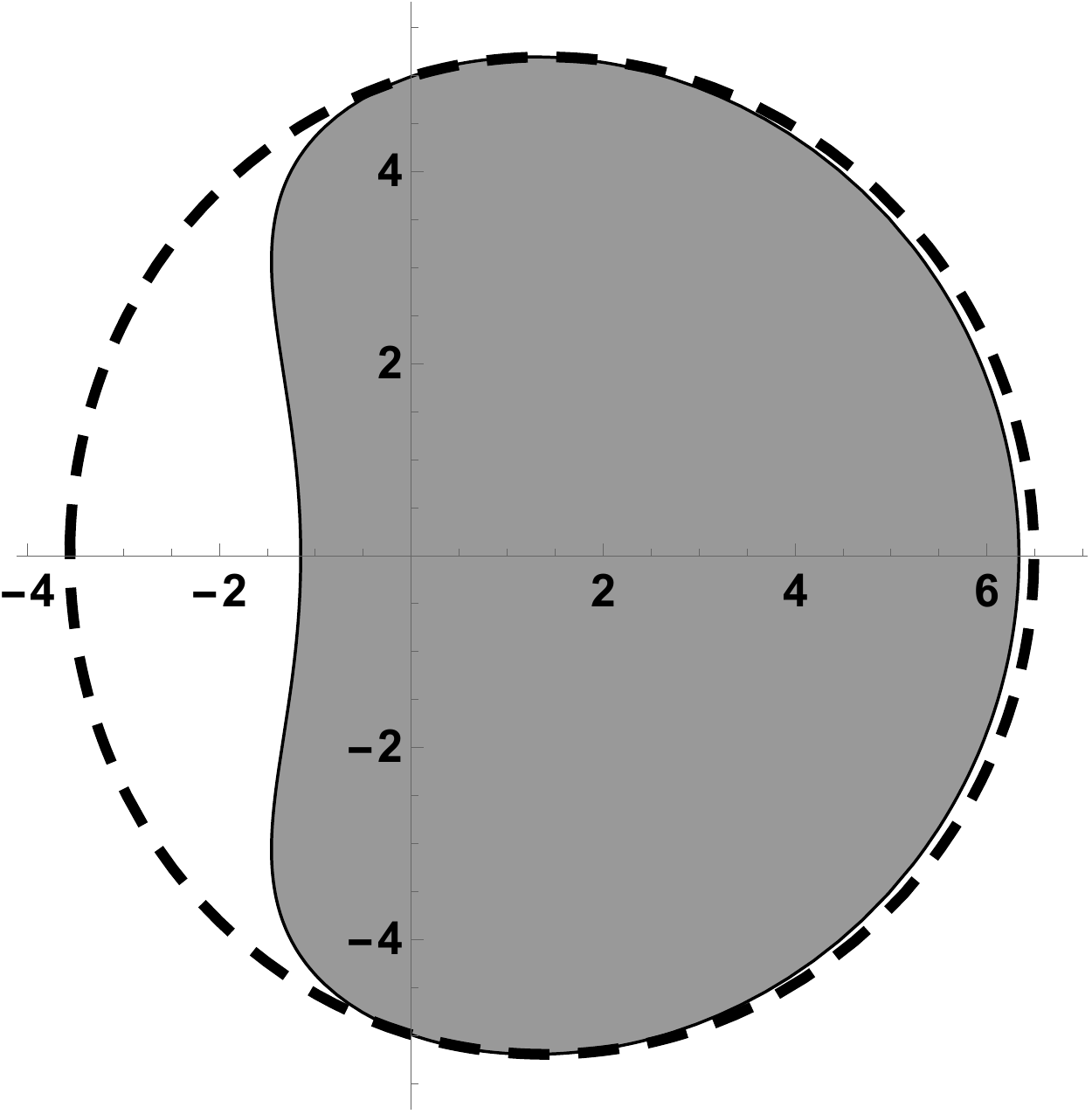}\label{re22p}}
\subfigure[Figure of deformed Kerr naked singularity with $\epsilon = 18$ and $a = 0.8$.]
{\includegraphics[width=55mm]{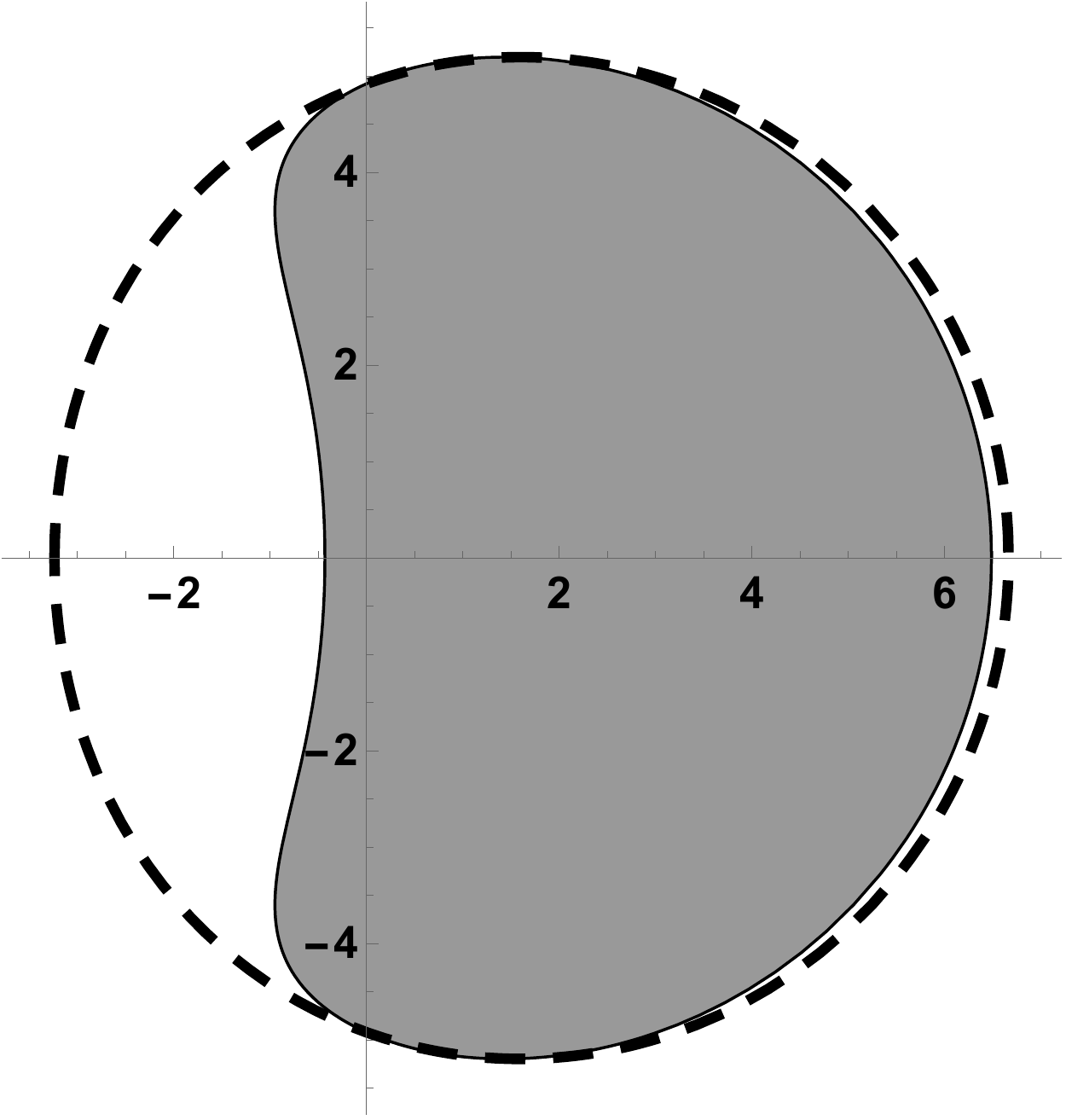}\label{re33p}}\\
 \caption{Figure shows the shape of the shadow cast by the deformed Kerr spacetime for various values of deformation parameter $\epsilon$ and spin parameter $a$. The details of this figure are discussed in the text.}
\label{figshadow}
\end{figure*}
%%%%%%%%%%%%%%%%%%%%%%%%%%%%%%%%%%%%%%%%%%%%%%%%%%%%%%%%%%%%%%%%%%%%%%%%%%%%%%%%%%%%%%

\begin{figure*}
\centering
{\includegraphics[width=140mm]{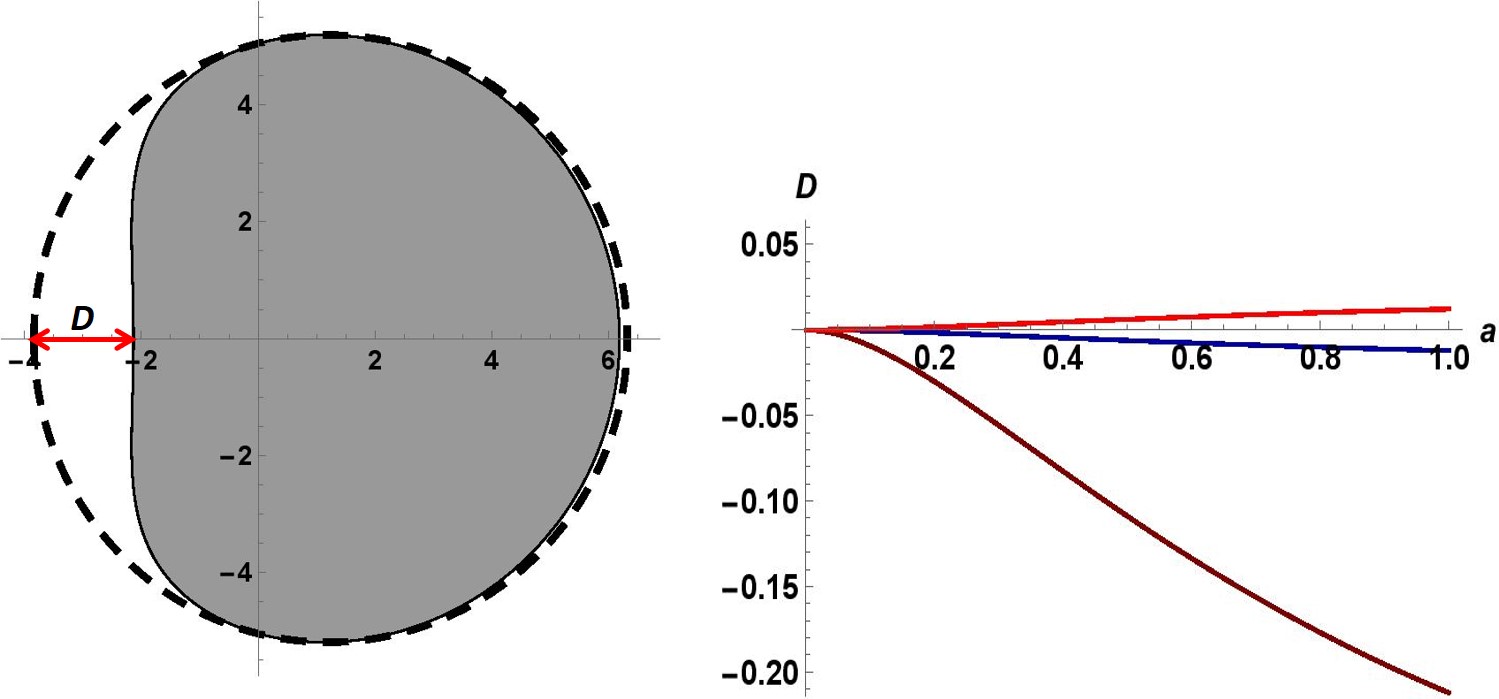}}
 \caption{Figure shows how the deformation of shadow shape ($D$) changes with the value of spin parameter $a$, where the red, blue and brown lines are corresponding to the deformed Kerr spacetime having $\epsilon=-1,1$ and $18$ respectively.}\label{D}
\end{figure*}
%%%%%%%%%%%%%%%%%%%%%%%%%%%%%%%%%%%%%%%%%%%%%%%%%%%%%%%%%%%%%%%%%%%%%%%%%%

\section{Shadow cast by deformed Kerr spacetime}
\label{shadowsec}
\subsection{Shadow size in deformed Kerr black hole}
In order to understand the nature of the geodesics in a spacetime, one can use the following Hamilton-Jacobi equation,
\begin{equation}
   2\frac{\delta S}{\delta \lambda} = - g^{\mu \nu}\frac{\delta S}{\delta x^{\mu}} \frac{\delta S}{\delta x^{\nu}} \label{eq1},
\end{equation}
where $\lambda$ is an affine parameter and S represent Hamilton's principal function or it can also be called as Jacobi action. Since we have the separable solution of the above differential equation, the Jacobi action can be written as,
\begin{equation}
   S = \frac{1}{2}\psi^2 \lambda - \mathcal{E}t + \mathcal{L} \phi + S_{r}(r) + S_{\theta}(\theta) \label{eq2},
\end{equation}
where $\mathcal{E}$ and $\mathcal{L}$ are the conserved energy and angular momentum (about the symmetry axis ) respectively and $\psi$ is the rest mass of the test particle. Therefore, for lightlike geodesic $\psi =0$. Using Eqs.~(\ref{eq1}), (\ref{eq2}), we can write down the following equations,
\begin{eqnarray}
\frac{(\Delta + h a^2 \sin^2{\theta})^2}{(1 + h)} \left(\frac{\delta S_{r}}{\delta r}\right)^2 &=& \mathcal{G}_1-\mathcal{G}_2\\
\frac{(\Delta + h a^2 \sin^2{\theta})}{\Delta} \left(\frac{\delta S_{\theta}}{\delta \theta}\right)^2 &=& \mathcal{K}- \mathcal{G}_3-\mathcal{G}_4\,\, ,
\end{eqnarray}
where $\mathcal{G}_1=\left[(r^2 + a^2)\mathcal{E} - a \mathcal{L} \right]^2$, $\mathcal{G}_2=\Delta \left[\mathcal{K} +(\mathcal{L} - a \mathcal{E})^2 + \delta_{1} r^2 \right]$, $\mathcal{G}_3=\left(a^2 \cos^2{\theta} + \frac{a^2 \sin^2{\theta} h}{\Delta} \Sigma \right) \delta_{1}$, $\mathcal{G}_4=(\mathcal{L}^2 \csc^2{\theta} - a^2 \mathcal{E}^2)\cos^2{\theta} - \frac{h \Sigma^2}{\Delta (1 + h)}$ and $\mathcal{K}$ is a separation constant which is also known as Carter constant. Using the above differential equations, one can write down the following radial part of equation of motion of photon, 
\begin{equation}
    \left(\Sigma \frac{dr}{d\lambda}\right)^2 -R(r) = 0\,\, ,
    \label{rmotion}
\end{equation}
where    
\begin{equation}
   R(r) = \left[(r^2 + a^2) - a \xi \right]^2 - \Delta \left[\eta + (\xi - a)^2 \right]\,\, ,
   \label{rpoten}
\end{equation}
where $\xi = \frac{\mathcal{L}}{\mathcal{E}}$,  $\eta = \frac{\mathcal{K}}{\mathcal{E}^2}$. The $R(r)$ can be defined as radial part of effective potential of lightlike geodesics. For the motion in $\theta$, one can define the following equation,
\begin{equation}
    \left(\Sigma \frac{d\theta}{d\lambda}\right)^2 -\Theta(\theta) = 0\,\, ,
    \label{thetamotion}
\end{equation}
where $\Theta(\theta)$ can be considered as $\theta $ part of effective potential which can be written as,
\begin{equation}
\Theta(\theta)=\mathcal{K}+a^2\mathcal{E}^2\cos^2\theta-\mathcal{L}^2\cot^2\theta\,\,.
\label{thetapoten}
\end{equation}
From Eqs.~(\ref{rpoten}), (\ref{thetapoten}), it can be verified that $R(r)\geq 0$ and $\Theta(\theta)\geq 0$ for the motion of photon.

In a general rotating spacetime, unstable circular orbits exist when the following conditions hold,
\begin{equation}
     R(r) = 0, \;   \frac{d R(r)}{dr}=0, \;    \frac{d^2 R(r)}{dr^2} \leq 0\,\,. \label{eq3}
\end{equation}
Using the above Eq.~(\ref{eq3}), we can get the critical impact parameters corresponding to the maxima of the $R(r)$ as,
\begin{equation}
     \xi = \frac{\left[r^2(r - 3M) + a^2 (r + M)\right]}{a (M - r)}
\end{equation}
\begin{equation}
    \eta = \frac{2\left[r^2(r^2 - 3M^2) + a^2 (r^2 + M^2)\right]}{(M - r)^2}
\end{equation}
From the above expressions of the critical impact parameters and the expression of $R(r)$ (Eq.~(\ref{rpoten})), it can be seen that the radial part of effective potential is independent of the value of deformation parameter $\epsilon$. Therefore, the nature of $R(r)$ is similar in both Kerr and deformed Kerr spacetime. As we know, in Kerr spacetime, for $a>1$, the Eq.~(\ref{eq3}) does not hold and therefore, there exist no shadow of the central naked singularity of the Kerr spacetime. Similarly, in deformed Kerr spacetime, when $a>1$, the central naked singularity cannot cast shadow.
%%%%%%%%%%%%%%%%%%%%%%%%%%%%%%%%%%%%%%%%%%%%%%%%%
\begin{figure*}
{\includegraphics[width=140mm]{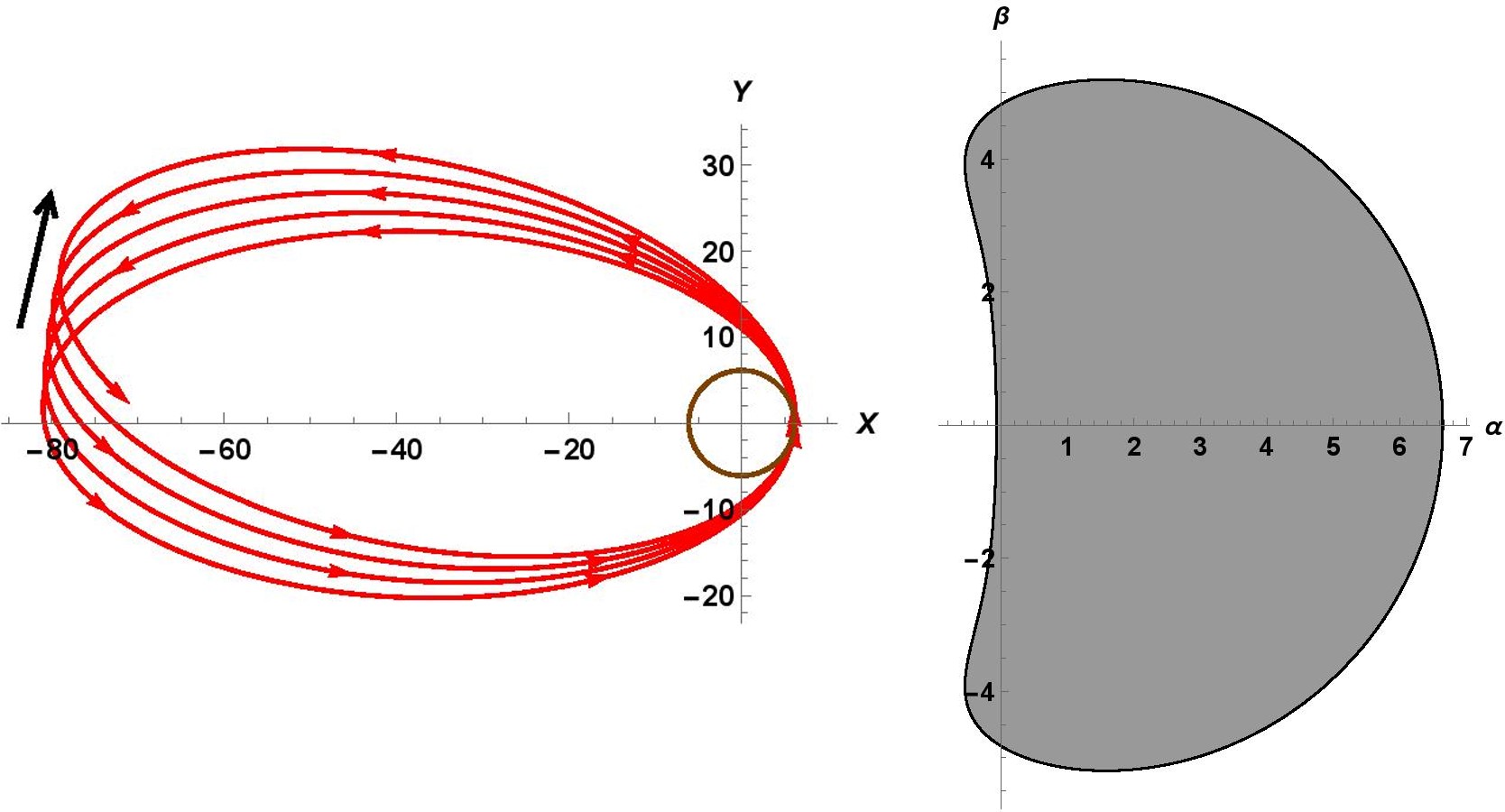}}
\caption{Figure shows the simultaneous occurence of the negative precession of bound timelike geodesics and the shadow in the deformed Kerr naked singularity spacetime, where we take $a=0.908584, \epsilon=18.68, e=0.988385$ and $L=3.20141$.}
\label{negpreshadow}
\end{figure*}

 In order to investigate the shadow shape in deformed Kerr spacetime, we need to derive the expressions of the celestial coordinates $\alpha$ and $\beta$, which can be written in the following way,
\begin{eqnarray}
    \alpha &=& \lim_{r_0\to\infty} \left(-r_0^2 \sin{\theta_{0}} \frac{d \phi}{dr}\right)\,\, ,\\
    \beta &=& \lim_{r_0\to\infty} \left(r_0^2 \frac{d \theta}{dr}\right)\,\, ,
\end{eqnarray}
where $r_0$ and $\theta_0$ are the coordinates of the asymptotic observer. From the definition of the celestial coordinates, one can write down the following expressions of celestial coordinates for deformed Kerr spacetime, 
\begin{eqnarray}
    \alpha &=& -  \frac{ - \Delta \xi}{(\Delta + a^2 h \sin^2{\theta_{0}})\sin{\theta_{0}}}\,\, ,\\
    \beta &=& \sqrt{\eta - a^2 \sin^2{\theta_{0}} - \xi^2 \csc^2{\theta_{0}}\,\, .}
\end{eqnarray}
From the above expressions of celestial coordinates, it can be verified that they are not independent of the deformation parameter $\epsilon$. Therefore, though the behaviour of $R(r)$ is similar for both Kerr and deformed Kerr spacetime, the shadow shapes of the both spacetimes are distinguishable from each other. In Fig.~(\ref{figshadow}), we show the shapes of the shadows cast by deformed Kerr spacetime for various values of $\epsilon$ and $a$, where the shaded region shows the shape of the shadow cast by deformed Kerr spacetime and the dashed lines show the shape of the shadow cast by Kerr spacetime.  Figs.~(\ref{re1n}),(\ref{re2n}) and (\ref{re3n}) show the shadow shape for $a=0.6, 0.7$ and $0.8$ respectively, where the $\epsilon=-1$. On the other hand, in Figs.~(\ref{re1p}),(\ref{re2p}) and (\ref{re3p}), we show the shadow shape for $a=0.6, 0.7$ and $0.8$ respectively, where the $\epsilon=1$. From those figures, one can see how the shadow shape changes when $\epsilon$ changes its signature. Figs.~(\ref{re11p}),(\ref{re22p}) and (\ref{re33p})also show the shadow shape of deformed Kerr spacetime for the same values of spin parameter, where we take $\epsilon=18$. As we discussed before, the central singularity in deformed Kerr spacetime cannot be naked when $\epsilon<0$ where $0\leq a\leq1$. Therefore, all the shadow shapes shown in Figs.~(\ref{re1n}),(\ref{re2n}) and (\ref{re3n}) are corresponding to the deformed Kerr black hole. However, though the Figs.~(\ref{re1p}),(\ref{re2p}) are showing the shape of the shadow cast by deformed Kerr black hole, Figs.~(\ref{re3p}),(\ref{re11p}),(\ref{re22p}) and (\ref{re33p}) are showing the shadow shape of the deformed Kerr naked singularity, which can be verified from Figs.~(\ref{pos3},\ref{pos3D}) and Fig.~(\ref{nakedcond}). Therefore, unlike Kerr naked singularity, the deformed Kerr naked singularity can cast shadow.

The amount of the deformation ($D$) of the shadow shape from the shadow shape of Kerr black hole can be defined as,
\begin{equation}
    \alpha_{dk} - \alpha_{k} = D\, ,
\end{equation}
where we consider $\beta=0$ axis to measure the deformation of the shadow shape and $\alpha_{dk}$ and $\alpha_{k}$ are celestial coordinates for deformed Kerr and Kerr spacetimes respectively.

One can derive following equation of $D$ for deformed Kerr spacetime,
\begin{equation}
     D = \frac{a \epsilon M^3 (r^2 (r - 3M) + a^2 (M + r))}{(M - r)(r^4 (r - 2M) + a^2 (\epsilon M^3 + r^3))}\,\, ,
\end{equation}
where
\begin{equation}
     r = \frac{\left(M + (2a^2 M - M^3 + 2 a M \sqrt{a^2 - M^2})^{1/3}\right)^2}{(2a^2 M - M^3 + 2 a M\sqrt{a^2 - M^2})^{1/3}}\,\, .
\end{equation}
Fig.~(\ref{D}) shows how the deformation of the shadow shape $D$ changes with spin parameter $a$ for various values of $\epsilon$. The left side figure in Fig.~(\ref{D}) is showing how one can define the deformation in the shadow shape. On the other hand, in the right side figure in Fig.~(\ref{D}), the red, blue and brown curves are showing how the deformation of the shape of the shadow changes with spin parameter $a$, when the $\epsilon= -1, 1$ and $18$ respectively.

In Fig.~(\ref{negpreshadow}), we show that in deformed Kerr spacetime, the negative precession of bound timelike orbits and the shadow both can exist simultaneously, where we take $a=0.908584, \epsilon=18.68, e=0.988385$ and $L=3.20141$. This is just one example. There are infinite number of such solutions where the shadow and negative precession occur simultaneously which is shown by the shaded region above the blue dashed line in Fig.~(\ref{enrgcon1}). However, this simultaneous occurrence of the shadow and the negative precession is present only in the deformed Kerr naked singularity spacetime which can also be seen from the Fig.~(\ref{enrgcon1}).    
\section{Conclusion}
\label{concludesec}
Followings are the important results what we get in this paper,
\begin{itemize}
\item In this paper, we show that unlike Kerr spacetime, in deformed Kerr spacetime \cite{Bam2020}, the negative precession of bound timelike geodesics is possible. We also show that the deformed Kerr black hole allows only the positive precession, however the deformed Kerr naked singularity allows both types of precession (i.e. positive and negative precession). 
%It is also shown that in deformed Schwarzschild spacetime, negative precession of bound timelike orbits is not possible in the allowed range of deformation parameter.
\item In this paper, we show that deformed Kerr naked singularity can cast shadow. In \cite{Bam2020}, we show that in Kerr spacetime, negative precession of timelike bound orbits is not possible, therefore there exist no possibilities of simultaneous existence of shadow and negative precession. In \cite{Dey:2020haf}, we show that JMN-1 and JNW (i.e. Janis-Newman-Winicour spacetime) naked singularity spacetimes do not allow the simultaneous occurrence of shadow and negative precession. Therefore, in that paper, we construct a spacetime structure which has a central naked singularity and which can cast shadow and allow negative precession of bound timelike orbits simultaneously. In this paper, we show that the shadow and the negative precession both can exist simultaneously in deformed Kerr naked singularity spacetime and that is very important novel feature of deformed Kerr spacetime.  
\end{itemize}
If there exists finite amount of deformation of Kerr spacetime then the effect of the deformation may be become very much significant near the singularity of deformed Kerr spacetime. One of the important and novel features of the deformed Kerr spacetime is that shadow and negative precession can exist simultaneously in that spacetime. As we know, UCLA galactic group, GRAVITY, SINFONI group and EHT collaboration are continuously observing the center of our milkyway galaxy Sgr-A*. Therefore, any evidence of such simultaneous occurrence of shadow and negative precession may imply the existence of deformed Kerr naked singularity at the center of our milkyway galaxy.

\newpage
\begin{widetext}
\section{Appendix}
\label{Appendix}
Here, we define some functions which are introduced in section (\ref{orbitsec}).

\subsection{Deformed Kerr spacetime}
\begin{eqnarray}
    &f_6(a,e,L,\epsilon,M,u) = 4 a^2 A^2 (-3 + 8 A) (a e - L)^3 M^2,\nonumber\\\nonumber\\
    &f_5(a,e,L,\epsilon,M,u) = 2 A M (-a e + L) (a^2 A (a^2 (2 A (-3 + 5 A) + (3 - 7 A) e^2) + a (3 - 10 A) e L + (-6 + 17 A) L^2)\nonumber \\\nonumber
    &\hspace*{3cm} + 12 (-a e + L)^2 M^2),\\\nonumber\\
    &f_4(a,e,L,\epsilon,M,u) = A (-3 a^4 A (e^2 + A (-2 + 4 A - 3 e^2)) L + 3 a^2 (1 - 3 A) A L^3 + 4 (-a e + L) (a^2 A (3 - 8 A + e^2)\nonumber \\\nonumber
    &\hspace*{3cm} + a (9 - 2 A) e L + (-9 + A) L^2) M^2),\nonumber\\\nonumber\\\nonumber
    &f_3(a,e,L,\epsilon,M,u) = -2 A M (a^3 e (A (-3 + 5 A) + (-3 + A) e^2) + a^2 ((6 - 17 A) A + 3 e^2) L\nonumber\\
    &\hspace*{3cm} + (-9 + 2 A) L^3 + 4 (-3 + 2 A) L M^2 + a e (-3 (-3 + A) L^2 + 4 (3 - 2 A) M^2)),\nonumber\\\nonumber\\\nonumber
    &f_2(a,e,L,\epsilon,M,u) = 3 a^2 A (e^2 + A (1 - 3 A + e^2)) L + (-3 + A) A L^3 + 4 (a e ((6 - 7 A) A \nonumber\\
    &\hspace*{3cm} + 6 (-1 + A) e^2) + (A (-9 + 7 A) - 6 (-1 + A) e^2) L) M^2,\nonumber\\\nonumber\\\nonumber
    &f_1(a,e,L,\epsilon,M,u) = 2 (a e (A (-3 + 5 A) + 2 (3 - 4 A) e^2) + ((9 - 8 A) A + 12 (-1 + A) e^2) L) M,\nonumber\\\nonumber\\\nonumber
    &f_0(a,e,L,\epsilon,M) = 3 (-1 + A) (A - 2 e^2) L,\nonumber
\end{eqnarray}
\begin{eqnarray}
    &g_4(a,e,L,\epsilon,p) = L (-2 + p)^2 (2 p^6 + p^3 (-2 + (3 - 6 e^2) p) \epsilon + (-4 + 3 p) \epsilon^2) + 2 a e (-2 p^6 (2 + p + (-1 + e^2) p^2) \nonumber\\
    &\hspace{3cm}+ p^3 (4 + p (-16 + e^2 (12 - 8 p) + 7 p)) \epsilon + (-2 + p) (-4 + 5 p) \epsilon^2),\nonumber\\
    \nonumber\\
    &g_2(a,e,L,\epsilon,p) = -6 a e L^2 (-2 + p) p^3 (2 (-3 + p) p^2 - \epsilon) + L^3 (-2 + p)^2 p^3 (-2 (-3 + p) p^2 + \epsilon) + 2 a^3 e (e^2 p^3 (2 + p) \nonumber\\
    &\hspace{3cm}(2 (-3 + p) p^2 - \epsilon) - (p^3 + \epsilon) (2 (-5 + p) p^3 + (-16 + 5 p) \epsilon)) + a^2 L (-(-2 + p) (p^3 + \epsilon)\nonumber \\
    &\hspace{3cm}(2 p^3 (-5 + 3 p) + (-16 + 9 p) \epsilon) + 3 e^2 p^3 (4 \epsilon + p^2 (24 + 2 p (-4 + (-1 + p) p) + \epsilon))),\nonumber\\\nonumber\\
    &g_0(a,e,L,\epsilon,p) = 6 a e L^2 p^2 (10 p^3 - 6 p^4 + 16 \epsilon - 9 p \epsilon) - L^3 (-2 + p) p^2 (2 p^3 (-5 + 3 p) + (-16 + 9 p) \epsilon) + 2 a^3 e (-2 (p^3 + \epsilon)\nonumber \\
    &\hspace{3cm}(2 p^3 + 5 \epsilon) + e^2 p^2 (2 p^3 (5 + 2 p) + (16 + 7 p) \epsilon)) + a^2 L (-2 (p^3 + \epsilon) (p^3 (-4 + 3 p) + 2 (-5 + 3 p) \epsilon)\nonumber \\
    &\hspace{3cm}+ 3 e^2 p^2 (2 p^3 (-10 + p + p^2) + (-32 + p (2 + 3 p)) \epsilon)),\nonumber
\end{eqnarray}
\begin{eqnarray*}
    s_6(a,e,L,\epsilon,p) = 3 L^2 (-2 + p)^4 \epsilon (p^6 (4 + (-1 + 2 e^2) p) + p^3 (2 + p - 7 e^2 p) \epsilon + 2 (-1 + p) \epsilon^2) + 4 a^2 e^2 ((e^2 (4 - 3 p) + 3 (-2 + p)) p^{11}\nonumber \\+ 2 p^6 (24 + p (6 p^2 - 9 (2 + p) + e^2 (12 + 4 p - 6 p^2))) \epsilon - 4 p^3 (-6 + p (-9 - 6 (-3 + p) p \\+ e^2 (21 + p (-28 + 9 p)))) \epsilon^2 + 3 (-2 + p) (4 + 5 (-2 + p) p) \epsilon^3) + 4 a e L (-2 + p) (3 (-2 + p)^2 \epsilon\nonumber \\(p^3 + \epsilon) (4 p^3 - 2 \epsilon + 3 p \epsilon) + e^2 p^4 (2 p^7 - 8 (-3 + p) p^3 \epsilon + (-84 + (98 - 27 p) p) \epsilon^2)),
\end{eqnarray*}
\begin{eqnarray*}
    s_4(a,e,L,\epsilon,p) = 8 a e L^3 (-2 + p)^3 p^3 ((-6 + p) p^5 + (12 - 7 p) p^2 \epsilon + \epsilon^2) + L^4 (-2 + p)^4 p^3 ((-6 + p) p^5 + (12 - 7 p) p^2 \epsilon + \epsilon^2)\nonumber \\+ 3 a^2 L^2 (-2 + p) ((-2 + p)^3 (p^3 + \epsilon)^3 + e^2 p^3 (-p^5 (-96 + p (64 - 8 p + p^3)) + (-4 + p) p^2 \\(48 + p(-40 + p(-2 + 7 p))) \epsilon - (-4 + p) (-4 + (-2 + p) p) \epsilon^2)) + 4 a^4 e^2 (12 (p^3 + \epsilon)^3\nonumber \\+ e^2 p^3 (4 (-6 + p) p^5 + p^2 (4 + 3 p) (12 + p (-16 + 3 p)) \epsilon + 2 (2 - 9 (-1 + p) p) \epsilon^2)) \\+ 4 a^3 e L (-3 (8 - 8 p + p^3) (p^3 + \epsilon)^3 + e^2 p^3 (p^5 (96+ p(-64 + 8 p + 3 p^3))\nonumber \\+ 8 p^2 (-24 + p (26 + p (2 + 3 (-3 + p) p))) \epsilon - 2 (8 + p (14 + 3 (-6 + p) p)) \epsilon^2)),
\end{eqnarray*}
\begin{eqnarray*}
    s_2(a,e,L,\epsilon,p) = -32 a e L^3 (-2 + p)^3 p^2 - 4 L^4 (-2 + p)^4 p^2 + 4 a^4 e^2 (-(-6 + p) p^3 + e^2 p^2 (-16 + p^2) - 10 (-3 + p) \epsilon)\nonumber \\
    + a^2 L^2 (2 e^2 p^2 (-192 + p (192 + p (-46 + p (-4 + 3 p)))) - (-2 + p) (2 p^3 (6 + p (-8 + 3 p)) + (60 +  7 p \nonumber\\
    (-10 + 3 p)) \epsilon)) + 8 a^3 e L (-30 \epsilon + p (e^2 p (32 + p (-16 + (-1 + p) p)) + 30 \epsilon - p (p (6 + (-6 + p) p) + 7 \epsilon))),
\end{eqnarray*}
\begin{eqnarray*}
    s_0(a,e,L,\epsilon,p) = -8 a e L^3 p^2 (p^3 (40 + 3 p (-16 + 5 p)) + 2 (56 + 9 p (-7 + 2 p)) \epsilon) - L^4 (-2 + p) p^2 (p^3 (40 + 3 p (-16 + 5 p))\nonumber \\
    + 2 (56 + 9 p (-7 + 2 p)) \epsilon) + 4 a^4 e^2 (2 p^5 (-3 p + e^2 (10 + 3 p)) + 7 p^2 (-6 p + e^2 (8 + 3 p)) \epsilon - 45 \epsilon^2)\nonumber \\
    + 3 a^2 L^2 (p^6 (-8 + (12 - 5 p) p) - 8 p^3 (7 + 3 (-3 + p) p) \epsilon - 2 (30 + p (-36 + 11 p)) \epsilon^2 + e^2 p^2 (p^3 (160 \\
    + p (-112 + p (2 + 5 p))) + 2 (-4 + p) (-56 + 3 p (7 + 2 p)) \epsilon)) + 4 a^3 e L (3 (4 - 3 p) p^6 + 6 (14 - 9 p) p^3 \epsilon\nonumber \\
    + 18 (5 - 3 p) \epsilon^2 + e^2 p^2 (p^3 (4 + p) (-20 + 9 p) + (-224 + p (28 + 27 p)) \epsilon)).
\end{eqnarray*}
\end{widetext}

\end{document}